# Doping Effect and Li-Ion Conduction Mechanism of $A$Li$_6$$X$O$_6$ ($A$ = K or Rb, and $X$ = Pentavalent): A First-Principles Study


Joohwi Lee[*] and Ryoji Asahi[†]

Toyota Central R&D Laboratories, Inc., Nagakute, Aichi 480-1192, Japan



* Correspondence: Joohwi Lee (j-lee@mosk.tytlabs.co.jp)

[†] Present address : Nagoya University, Nagoya, Aichi 464−8603, Japan





**ABSTRACT**

Recent theoretical and experimental evaluations demonstrated that $KLi_6TaO_6$ is a good Li-ion conductor. In this study, the energetics and detailed mechanism of Li-ion migration, relevant to the point defects of $KLi_6TaO_6$, were analyzed by first-principles calculations. Defect formation energy analysis suggested that it has limited chemical potential conditions for attaining Li-excess conditions through doping (substituting tetravalent elements for Ta). The formation of other native defects, such as Li vacancies, hinders the stabilization of the dopant and compensates for the interstitial Li. When the doping is successful, the interactions between the coexisting dopant and interstitial Li can increase the migration energy barrier of the interstitial Li. This phenomenon limits the factors responsible for achieving high Li-ion conductivity in this material. Based on the results of the investigations on $KLi_6TaO_6$, isostructural materials of the form $ALi_6XO_6$, with various combinations of constituent elements $A$ and $X$, were each screened on the basis of high stability and low Li-ion migration energy. Twelve structures of the form ($A$ = K or Rb)$Li_6XO_6$ were suggested, of which $X$ was pentavalent. They also exhibited limited chemical potential conditions for achieving Li-excess conditions through doping. Combinations of the suggested isostructural oxides and dopants were identified to reduce the interactions between interstitial Li and dopant. Some isostructural oxides were doped using Sn and were evaluated using first-principles molecular dynamics; their Li-ion conductivities at room temperature were found to be comparable with those of garnet-type Li-ion conductors.






# 1. Introduction

The development of a solid-state electrolyte with high Li-ion conductivity ($\sigma_{Li}$) is essential to fabricate an all-solid-state Li-ion battery.[1] Many solid sulfides have been developed because of their high $\sigma_{Li}$ and simple fabrication methods. The highest reported $\sigma_{Li}$ at room temperature ($\sigma_{Li}$ at 300 K) for a solid-state electrolyte is 25 mS/cm in $Li_{9.54}Si_{1.74}P_{1.44}S_{11.7}Cl_{0.3}$,[2] which is a derivative of $Li_{10}GeP_2S_{12}$ (LGPS).[3,4] Solid oxides have also been developed as stable Li-ion conductors; however, when compared with sulfides, Li-ion conducting oxides, such as garnet-type[5-9] and perovskite[10-12] oxides have $\sigma_{Li}$ values that are 1–2 orders of magnitude lower. Because typical Li-ion conducting oxides have more than three constituent elements, there are still undiscovered Li-ion conducting oxides considering the high number of possible combinations of constituent elements.

Recently, Suzuki *et al.*[13] suggested $KLi_6TaO_6$, with a crystal structure that is different from that of garnet-type or perovskite oxide, as a potential Li-ion conductor; this was based on screening by a corrugation descriptor[14] that efficiently estimates the Li migration energy of materials using first-principles calculations. $KLi_6TaO_6$ yielded a low migration energy barrier ($E_m$) and activation energy ($E_a$) of approximately 0.1 eV, which were evaluated using nudged elastic band (NEB) and first-principles molecular dynamics (FPMD) under Li-excess conditions, respectively. In addition, the estimated $\sigma_{Li}$ at 300 K was in the order of $10^{-2}$ S/cm. However, experimental evaluation of the synthesized material did not result in such a high $\sigma_{Li}$. Sn-doped $KLi_6TaO_6$, which was intended to achieve Li-excess conditions, exhibited a $\sigma_{Li}$ at 300 K that was 3–4 orders of magnitude lower than the theoretical value. One possible reason for this lowered value is that the Li-excess condition, which is essential for achieving a low $E_a$, was not realized.

Meanwhile, the corrugation descriptor determined that the other oxides that are isostructural with $KLi_6TaO_6$, such as $KLi_6BiO_6$ and $KLi_6IrO_6$, were also good Li-ion conductors.[14] In addition, FPMD predicted Sn-doped $KLi_6BiO_6$ to have a high $\sigma_{Li}$ at 300 K ($6\times10^{-3}$ S/cm).[15,16] Therefore, these findings indicate that it is worthwhile to investigate the potential of other isostructural oxides with $KLi_6TaO_6$ as Li-ion conductors. The material exploration of Li-ion conductors *via* total or partial exchange of constituent elements is a good method. Particular crystal structure families, such as garnet-type,[17] perovskite,[18] and argyrodite structures[19], with



various combinations of constituent elements, commonly have high $\sigma_{Li}$ values because they have particular geometric features that allow for high Li-ion mobility.

In this study, the detailed mechanism of Li-ion migration and the reason for the significantly lower experimentally obtained $\sigma_{Li}$ value for $KLi_6TaO_6$ than the theoretical value are discussed. Primary material screening of 58 types of oxides, which are isostructural with $KLi_6TaO_6$, was performed on the basis of high stability and low $E_m$. Based on the screening results, 12 oxides (including $KLi_6TaO_6$) were selected. These were further invesitaged to determine their ability to overcome the shortcomings of $KLi_6TaO_6$ and realize high $\sigma_{Li}$.

## 2. Method

### 2.1. Crystal structure

Figure 1 shows the crystal structure of $A Li_6 X O_6$ (space group $R$-$3m$), a general formula derived from $KLi_6TaO_6$. To satisfy charge neutrality, the sum of the valences of $A$ and $X$ must be 6. The conventional cell with a hexagonal structure has the formula $A_3Li_{18}X_3O_{18}$. The cations Li, $X$, and $A$ form chemical bonds with 4, 6, and 12 neighboring O atoms, respectively. Along the $c$-axis of the hexagonal cell, $LiO_4$ tetrahedra are connected, forming hexagons with large $A$ atoms at the center. Along the $ab$-plane view, it can be confirmed that Li-containing layers were alternately formed with the other O-containing layers.

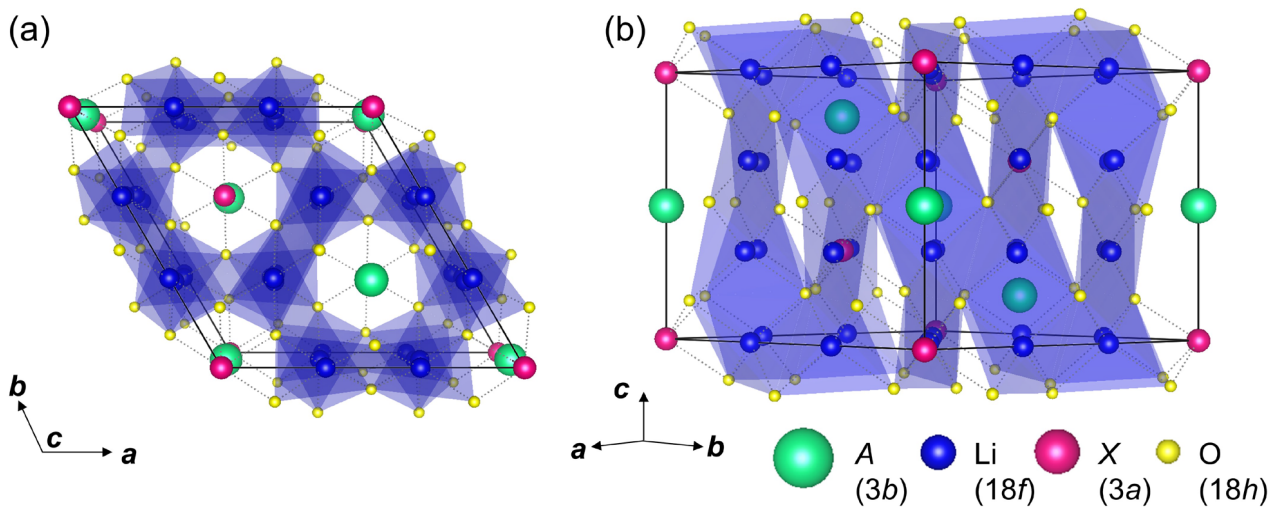

Figure 1. Crystal structure of $A Li_6 X O_6$ (space group of $R$-$3m$) in a conventional cell (with the formula $A_3Li_{18}X_3O_{18}$ in a hexagonal structure) with a perspective view along (a) top and (b) side directions. Solid and dotted lines indicate unit cell boundary and chemical bond, respectively. Connections of $LiO_4$ tetrahedra are also displayed to show the distributions of Li sites.



## 2.2. Exploration of isostructural Li-ion conductors

Additional candidates, with the same crystal structure as $KLi_6TaO_6$, were investigated as Li-ion conductors. For the $ALi_6XO_6$ structure shown in Figure 1, the $3b$ site (for the $A$ atom) and $3a$ site (for the $X$ atom) were occupied by various cations with different valences, while maintaining charge neutrality. When the $3b$ site was occupied by Ag, Na, K, or Rb (all monovalent cations), the $3a$ site was occupied by V, Nb, Ta, As, Sb, or Bi (all pentavalent cations). When the $3b$ site was occupied by Ca, Sr, Ba, or Cd (all divalent cations), the $3a$ site was occupied by Si, Ge, Sn, Ti, Zr, or Hf (all tetravalent cations). Finally, when the $3b$ site was occupied by La or Nd (both trivalent cations), the $3a$ site was occupied by Sc, Y, Al, Ga, or In (all trivalent cations). In total, 58 combinations were prepared.

Three screening conditions were used to identify the Li-ion oxides with good conductivities: $\Delta H_f$ ($ALi_6XO_6$) $< \Sigma\{\Delta H_f$ (binary oxides)$\}$, $w_{ph}^2 \geq 0$, and $E_m < 0.15$ eV. $\Delta H_f$ is the relative formation energy of the oxide compared with the energy summation of the most stable phases of each element (the most stable metals for cations and oxygen gas molecule for O). $w_{ph}$ denotes the eigenvalues of phonons; therefore, the second condition ensures dynamical stability. $E_m$ was obtained from the migration path shown in Figure 2a under Li-excess conditions, as described in the next subsection. The reasons for selecting such conditions are explained in Section 3.2.



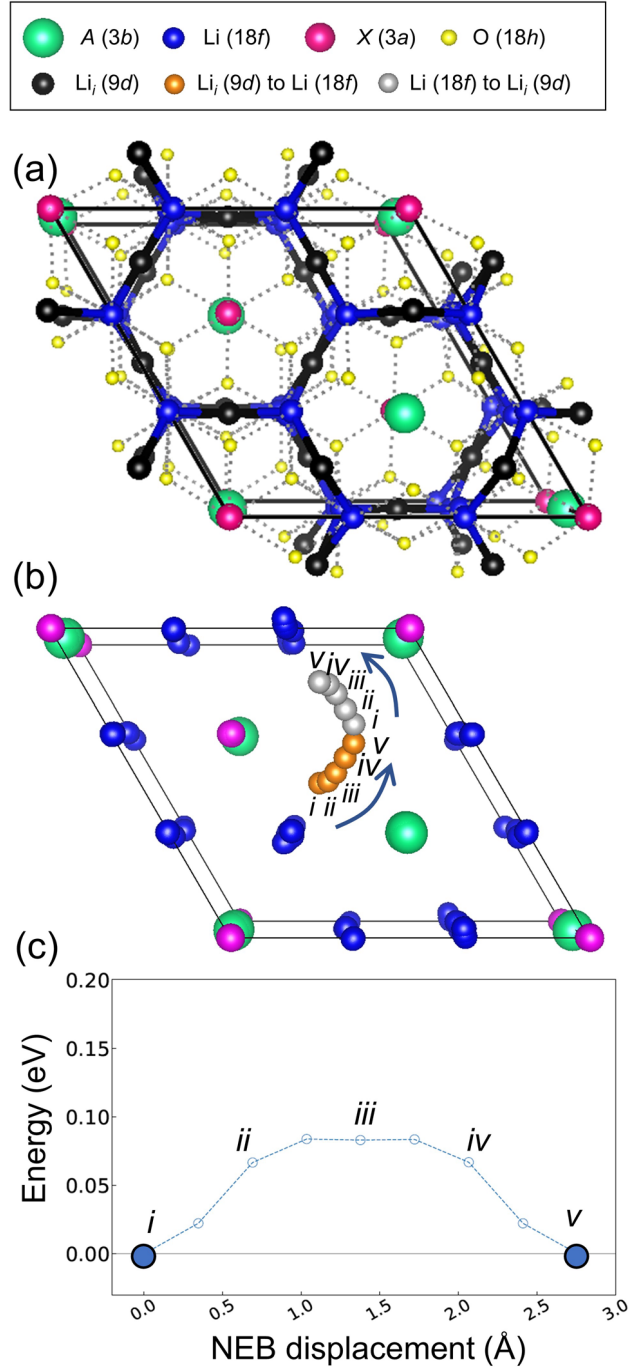

Figure 2. (a) Perspective view of the locations of stable Li interstitial sites (9$d$) and their migration paths in the $A$Li$_6$$X$O$_6$ structure. The identical migration paths with a kick-out mechanism are indicated using thick blue-black connected lines between 9$d$ interstitial and 18$f$ lattice sites. The chemical bonds between the cations and O atoms are displayed as gray dotted lines. In the actual calculation, one Li atom was added to achieve the Li-excess condition using $A$Li$_6$$X$O$_6$. (b) Trajectory of the moving atoms in $A$Li$_6$$X$O$_6$ under Li-excess conditions determined using NEB calculations. Orange atoms indicate the trajectory of a Li atom moving from an interstitial site to a lattice site, gray atoms indicate the trajectory of a different Li atom moving from a lattice site to an interstitial site. Note that the O and Li atoms at the lattice sites, which cover the migration path, are removed for better visualization. (c) NEB profile of KLi$_6$TaO$_6$ under Li-excess conditions (using a conventional cell with the formula K$_3$Li$_{19}$Ta$_3$O$_{18}$). The closed and open circles indicate initial and final states where the atomic coordinates were fixed and optimized during the NEB calculations, respectively.

**2.3. First-principles calculations**

All first-principles calculations were performed using the projector augmented wave (PAW)[20,21] method implemented in the Vienna *Ab initio* Simulation Package (VASP).[22,23] We used the exchange-correlation function of the generalized gradient approximation (GGA) parameterized in the Perdew–Burke–Ernzerhof (PBE) form.[24]

The optimization of a primitive cell (with 14 atoms) in a trigonal structure was performed until the interatomic force on each atom was reduced to within 0.005 eV/Å, with the associated changes in lattice constants and atomic coordinates. As an input structure, the crystal structure of $KLi_6TaO_6$[25] was used to optimize the crystal structures of the 58 types of $ALi_6XO_6$. The transformations between the primitive and conventional cells (hexagonal cell with the formula $A_3Li_{18}X_3O_{18}$, shown in Figure 1) were performed by SPGLIB,[26] implemented in the PHONOPY code.[27] A cutoff energy of 520 eV was used and the Brillouin zone was sampled using Γ-centered 4 × 4 × 4 meshes with integration based on Gaussian smearing (width: 0.1 eV). Electronic spin polarizations were activated, except in the case where the up- and down-spins compensated for each other. The bandgaps ($E_g$) were calculated using a screened hybrid density functional parameterized in Heyd–Scuseria–Ernzerhof (HSE06)[28] and GGA.

2.3.1. Defect formation energy

The defect formation energies ($E_f$) for point defects ($α^q$) with a charge $q$ in the $ALi_6XO_6$ structure were calculated using a supercell (2 × 2 × 2 primitive cell size) as follows:[29-32]

$$E_f^q = E(α^q) - E(\text{host}) - \sum_i \Delta n_i \mu_i^* + q(E_{VBM} + E_F) + \Delta E_{LZ}, \qquad (1)$$

where $E(α^q)$ and $E(\text{host})$ are the energy of supercells with and without point defects, respectively. $\Delta n_i$ is the change in the number of atoms of element $i$, which was either added ($\Delta n_i > 0$) or removed ($\Delta n_i < 0$). $\mu_i^*$ is the atomic chemical potential of element $i$. $E_{VBM}$ is the eigenvalue of valence band maximum (VBM). $E_F$ is the Fermi level, a variable referenced to the host VBM. $\Delta E_{LZ}$ is the correction term suggested by Lany and Zunger[33,34] to compensate for image charge interactions between supercells for charged point defects.

In this study, the value of $\mu_O^*$ was fixed as the half–energy of $O_2$ − 1.2 eV (= −6.13 eV), which corresponds to a synthesis temperature of 1073 K.[13] An additional pressure-related term was not considered because the calcination of the sample was performed in an oxygen gas stream. The dependence of $\mu_O$ on temperature and pressure was reported by Reuter et al.[35]

For a cation $\chi$, $\mu_\chi^*$ includes the total energy term of the most stable metal. Therefore, $\mu_\chi^* = E$ (the most stable metal for $\chi$) + $\mu_\chi$, where $\mu_\chi$ is the chemical potential referenced to the energy of the metal. Note that for $\mu_O^*$, the total energy term of the half–energy of $O_2$ was inlcuded, and $\mu_O$ was fixed at −1.2 eV. Equation 2 indicates the stability condition for the host material $A$Li$_6$XO$_6$.

$$\mu_A + 6\mu_{Li} + \mu_X + 6\mu_O = \Delta H_f(A\text{Li}_6XO_6). \tag{2}$$

Each chemical potential in the host material should satisfy the condition depicted in equation 3.

$$\mu_A, \mu_{Li}, \mu_X, \text{and } \mu_O \leq 0. \tag{3}$$

In addition, $A$Li$_6$XO$_6$ is more stable than a mixture of binary oxides. Therefore, the conditions represented by equation 4 should be satisfied.

$$x\mu_\chi + y\mu_O \leq \Delta H_f(\chi_x O_y), \tag{4}$$

where $\chi_x O_y$ is the binary oxide including cation $\chi$ ($A$, Li, and $X$ for $A$Li$_6$XO$_6$). Here, binary oxides with the usual oxidation numbers of cations were used, as mentioned in Section 2.2. For example, for KLi$_6$TaO$_6$, K$_2$O, Li$_2$O, and Ta$_2$O$_5$ were used. The chemical potential of the dopant, $M$ ($\mu_M^*$), was determined by the energy difference between the corresponding binary oxide ($MO_2$) and $\mu_O^*$ ($M$-rich condition). Tetravalent elements (Ge, Ti, Zr, Sn, Pb, Hf, and Ce) were used as dopants.

For determining the conditions of $\mu$, the concentration of each defect ($c(\alpha^q)$) is calculated by using equation 5:

$$c(\alpha^q) = c_0 \exp\left(\frac{-E_f^q}{k_B T}\right), \tag{5}$$

where $c_0$ is the concentration without any defects, $k_B$ is the Boltzmann's constant, $T$ is the temperature, and $E_f^q$ is obtained by using equation 1. In addition, the charged defect concentration and electronic carrier densities are constrained by the charge neutrality condition as obtained from equation 6:

$$p - n + \Sigma(q \cdot c(\alpha^q)) = 0, \tag{6}$$

where $p$ and $n$ are hole density in VBM and electron density in conduction band minimum (CBM), respectively, which are obtained by integrating the electronic density of states in the perfect crystal multiplied by Fermi-Dirac distribution from $-\infty$ to VBM and from CBM to $\infty$, respectively. The equilibrium $E_F$ ($E_F^{eq}$) at a given temperature can be determined simultaneously using the charge neutrality condition.

Interstitial Li ($Li_i^+$) and vacancy ($\square_{Li}^-$) were employed as native Li point defects. Because Li occupies identical sites, the site for $\square_{Li}^-$ is the same as that of the Li lattice site. $Li_i^+$ occupies the 9$d$ site, which is at the center of four Li 18$f$ sites, as shown in Figure 2a. The multiplicity of the Wyckoff position corresponds to that of a conventional cell. In addition, interstitials and vacancies of other elements were employed as native point defects.

Calculations using a supercell including point defects were performed until the interatomic forces on each atom were reduced to within 0.02 eV/Å. The lattice constants were fixed and the Brillouin zone was sampled using Γ-centered 2 × 2 × 2 meshes with integration based on Gaussian smearing (width: 0.1 eV). For calculations of charged defects, a neutralizing background was applied to the supercell. When a dopant and an interstitial Li coexisted in the supercell, compensation for the background charge was not employed because charge neutrality was satisfied. Among the various configurations, the lowest energy was used for $E(\alpha^q)$.

2.3.2 Nudged elastic band calculations

$E_m$ was calculated using the climbing image-nudged elastic band (CI-NEB) method,[36,37] with seven intermediate images, using a conventional cell or a supercell made of a 2 × 2 × 2 primitive cell. The CI-NEB calculations were performed until the forces decreased below 0.03 eV/Å, with a spring constant of 5 eV/Å² between the images. The $k$-space was sampled using Γ-centered 2 × 2 × 2 meshes. The $E_m$ values for the native Li defects were obtained with and without dopants. When dopants were not included, we used singly charged $\square_{Li}^-$ and $Li_i^+$ to obtain $E_m$, with a neutralizing background applied to the supercell. The migration path for $Li_i^+$ based on the kick-out mechanism (in other words, two-ion cooperative migration)[38] is shown in Figure 2a. The $E_m$ value was obtained as the difference between the highest and lowest energies of the CI-NEB profile.

2.3.3 First-principles molecular dynamics

To investigate the Li-ion diffusivity ($D_{Li}$), we performed FPMD using a supercell of the same size as that in the computations for $E_m$, including one $Li_i^+$ randomly occupying the interstitial site. For computational efficiency, the cut-off energy was slightly decreased to 500 eV, the spin polarizations were not activated, and only the $\Gamma$-point was used for $k$-space sampling. To accelerate the diffusion of Li atoms, a high $T$ range of 673–1273 K was chosen. We confirmed that the mean square displacement (MSD) of atoms other than Li was not greater than 1 Å$^2$ at such high $T$ values during FPMD (see Supplementary Figure S7). Therefore, $D_{Li}$ was obtained from the investigated $A$Li$_6X$O$_6$ structures without a phase transition.

The diffusion coefficient ($D$) of each chemical element at high $T$ can be obtained from the MSD after a long simulation time $t$ using the following equation:[39,40]

$$D = \lim_{t \to \infty} \frac{\text{MSD}}{6t}. \tag{7}$$

MSD ($t$, $t_0$) can be obtained using the following equation:

$$\text{MSD}(t, t_0) = \frac{1}{n}\sum_i |r_i(t + t_0) - r_i(t_0)|^2, \tag{8}$$

where $n$ is the number of atoms of each chemical element, and $r_i(t_0)$ and $r_i(t + t_0)$ are the positions of the $i^{th}$ atom at multiple time origins $t_0$ and new time $t + t_0$, respectively. The MSD was smoothed by averaging MSD ($t$, $t_0$) to improve the statistics. An interval of 1 fs was used for each atomic movement cycle. $T$ was elevated from 573 K to the target $T$ using 1000 cycles. FPMD was then performed based on the Nosé thermostat[41,42] with a canonical ensemble for 60000 cycles with sufficient movements of Li atoms. MSD was obtained by taking the average over a time from 3 to 59 ps. Three trials of FPMD were performed at each $T$ to obtain the average $D$.

$\sigma_{Li}$ can be estimated using $D_{Li}$ at a lower $T$ based on the following procedure. The $D_{Li}(T)$ at a certain $T$ can be obtained using the Arrhenius equation:

$$D_{Li}(T) = D_{Li,0} \exp\left(\frac{-E_a}{k_B T}\right), \tag{9}$$

where $D_{Li,0}$ is the pre-exponential factor, $E_a$ is the activation energy. $E_a$ was calculated from the gradient of the ln $D_{Li}$ vs. $1/T$ plot [or ln($T \cdot \sigma_{Li}$) vs. $1/T$ plot]. The $\sigma_{Li}(T)$ at the dilute level can be obtained from $D_{Li}(T)$ using the Nernst–Einstein equation as follows:[43,44]

$$T \cdot \sigma_{Li}(T) = \frac{q^2 c}{k_B} D_{Li}(T) = \frac{q^2 c}{k_B} D_{Li,0} \exp\left(\frac{-E_a}{k_B T}\right), \tag{10}$$

where $q$ is the unit charge ($1.6 \times 10^{-19}$ C) and $c$ is the ionic concentration of the Li atoms in the computational cell.

2.3.4 Phonon calculations

The dielectric constants for the correction term of the image charge interaction were computed using primitive cells with the density functional perturbation theory.[45]

The phonon calculations for dynamical stability were performed under the harmonic approximation of the lattice Hamiltonian using the PHONOPY code.[27] The calculations were performed using the supercell of the same size as that in the computations for $E_m$. The force constants of the oxides were calculated using the finite displacement method. The phonon frequencies were calculated from a dynamical matrix built from the force constants.

## 3. Results and Discussion
### 3.1 Summary of the properties of KLi$_6$TaO$_6$

In this sub-section, the characteristics of KLi$_6$TaO$_6$ reported in our previous studies[13,14] are summarized along with additional detailed analyses and explanations.

First, we have developed a corrugation descriptor that can quickly evaluate the energy barriers for Li atoms in Li-containing oxides.[14] It is a topological analysis based on electrostatic potential and repulsive energies between Li atoms and anions. KLi$_6$TaO$_6$ and KLi$_6$BiO$_6$ in the same crystal structure exhibited low corrugation barriers for Li atoms among various oxides in the Inorganic Crystal Structure Database (ICSD).[46] Therefore, these oxides are expected to have adequate geometric features to provide a low energy barrier for Li atoms. However, the main charge carriers relevant to Li atoms have not yet been investigated.

Then, we have focused on KLi$_6$TaO$_6$, which has rarely been investigated as a Li-ion conductor. This oxide was studied both theoretically and experimentally,[13] and the migration of Li atoms under Li-excess condition was investigated. The details of the migration mechanism of Li atoms are described herein. Figure 2a shows the connection between the Li lattice and interstitial sites in the $A$Li$_6$$X$O$_6$ structure. In a pure structure, Li atoms occupy the 18$f$ lattice sites. When a Li-excess condition is formed, a Li atom can occupy an interstitial 9$d$ site located among Li lattice sites. The interstitial and lattice sites are connected in three dimensions.



Li atoms can migrate *via* a kick-out mechanism as shown in Figure 2b. When the Li-excess condition is formed, an excess Li atom occupies the interstitial site, and its neighboring Li atoms at the lattice sites can be stressed (state *i*). If the Li atom at the interstitial site moves toward one neighboring lattice site, the Li atom at the lattice site must move toward another interstitial site (state *ii*). This intermediate state appears to be made of two Li$_i$ and one $\square_{Li}$ (state *iii*). Finally, the movement of Li atoms progresses, the Li atom coming from the interstitial site occupies the lattice site, and the Li atom kicked out from the lattice site moves to an adjacent interstitial site (state *v*). Therefore, the types of moving Li atoms and occupied interstitial sites changed for each movement.

Figure 2c shows the NEB profiles of the kick-out mechanism in KLi$_6$TaO$_6$ under Li-excess conditions (using a conventional cell with the formula K$_3$Li$_{19}$Ta$_3$O$_{18}$). $E_m$ exhibited a low value of 0.084 eV. Note that a similar value (0.086 eV) was obtained using a larger 2 × 2 × 2 primitive cell with the formula K$_8$Li$_{49}$Ta$_8$O$_{48}$. The NEB profile shows a symmetric double peak, but it is close to a plateau. It can also be seen that the intermediate state, state *iii*, on the plateau converges well without changing to another state when additional density functional theory calculations are performed by relaxing the internal coordinates. Therefore, this could be a dynamically stable state, and two Li$_i^+$ and one $\square_{Li}^-$ could coexist in a metastable state. However, under stoichiometric conditions, the atomic coordinates with a pair of Li$_i^+$ and $\square_{Li}^-$, which were intentionally prepared, were not dynamically stable and soon relaxed to those of pure KLi$_6$TaO$_6$. This suggests that a non-stoichiometric Li-excess condition is essential for stabilizing the occupancy of Li atoms at the interstitial sites.

In this study, for a comparison, $E_m$ values under Li-deficient conditions were also investigated. There are three likely migration paths for $\square_{Li}^-$. As shown in Supplementary Figure S1a, in the *ab*-plane direction, there are two migration paths alternately linked; along the *c*-axis direction, there is one migration path. Supplementary Figure S1b shows the NEB profiles of the native Li defects in KLi$_6$TaO$_6$. The $E_m$ along the *c*-axis is 0.25 eV, but those in the *ab*-plane direction are significantly larger (0.40 and 0.94 eV). These values for $\square_{Li}^-$ are much larger than those for Li$_i^+$. In addition, for the connection of migration paths in three dimensions, three types of $E_m$ of $\square_{Li}^-$ should be considered together. Therefore, these results provide a rationale for the formation of a Li-excess condition that makes $A$Li$_6$$X$O$_6$ a good Li-ion conductor.

In the previous study,[13] FPMD calculations were performed at multiple temperatures to estimate $\sigma_{Li}$ at 300 K. As a result, under a 2.1% Li-excess condition in KLi$_6$TaO$_6$ (K$_8$Li$_{49}$Ta$_8$O$_{48}$), a high value of 3×10$^{-2}$

S/cm was obtained for $\sigma_{Li}$ at 300 K with an $E_a$ of 0.11 eV. The trajectory of the Li atoms is in agreement with the migration path predicted by the NEB calculations. However, in an experimental validation, Sn-doped KLi$_6$TaO$_6$, which was intended to form a Li-excess condition, exhibited a $\sigma_{Li}$ of 6×10$^{-6}$ S/cm at 300 K and a larger $E_a$ of 0.37 eV. The values were closer to those obtained under Li-deficient conditions ($\sigma_{Li}$ of 7×10$^{-5}$ S/cm at 300 K and $E_a$ of 0.29 eV) than those obtained under Li-excess conditions based on FPMD calculations. In addition, Rietveld analysis exhibited that significant $\square_{Li}^-$ was indicated and sufficient amount of Sn was not substituted for Ta site. The other synthesized samples with dopants such as Ti, Zr, Hf, and Fe exhibited lower $\sigma_{Li}$ and higher $E_a$ than those of Sn-doped samples. Accordingly, it is considered that forming Li-excess condition by doping was not successful.

In this study, the dependence of $E_f$ of various defects on $\mu$ conditions in KLi$_6$TaO$_6$ was investigated at 1073 K (synthesis condition).[13] Sn was selected as a substituting dopant for Ta sites to form a negatively charged defect, Sn$_{Ta}^-$. Figure 3a shows the stability conditions for KLi$_6$TaO$_6$ under different $\mu$ conditions against binary oxides imposed by conditions (3) and (4). $\mu_O$ was fixed at −1.2 eV, which corresponds to the pressure and temperature during the synthesis. The allowed stability condition can be displayed by two variables of $\mu$ for cations. $\mu$ of the other cation is dependent by Eq. 2. It is noteworthy that the energies of all other competing compounds including K, Li, Ta, or O had to be compared with that of KLi$_6$TaO$_6$; thus, the allowed stability condition might be more limited than that determined by conditions (3) and (4). However, to avoid complexity with hundreds of competing compounds, the binary oxide constraint was considered as a necessary condition. Alternatively, the existence of a negative $E_f$ at $E_F^{eq}$ for any defects is considered that the host material is unstable because the defect concentration is higher than that of the sites. A "richer" condition for an element $\chi$ indicates a higher $\mu_\chi$. As a $\chi$-richer condition forms (higher $\mu_\chi$), $E_f$ of $\chi_i$ decreases, whereas $E_f$ of $\square_\chi$ increases. Conversely, as a $\chi$-poorer condition forms (lower $\mu_\chi$), $E_f$ of $\chi_i$ increases, whereas $E_f$ of $\square_\chi$ decreases.

Figure 3b shows the change in $E_f$ of various defects as a variable of $E_F$ under the $\mu$ condition *I*, which indicates K-rich, Ta-rich, and Li-poor conditions. At $E_F^{eq}$, the lowest $E_f$ values for positively and negatively charged defects were given by Li$_i^+$ and $\square_{Li}^-$, respectively. Because $E_f$ of $\square_{Li}^-$ was lower than that of Li$_i^+$, Li$_i^+$ is expected to recombine with excess $\square_{Li}^-$. $E_f$ of Sn$_{Ta}^-$ was much higher than that of $\square_{Li}^-$. Therefore, in this $\mu$ condition, achieving Li-excess conditions by doping is hindered by the active formation of $\square_{Li}^-$.

Figure 3c shows the results for the $\mu$ condition *II*, which indicates Li-rich, Ta-rich, and K-poor conditions. Under this condition, $E_f$ of $\square_K^-$ was significantly low, exhibiting negative values for the entire $E_F$. $E_F^{eq}$ was not located within $E_g$ and is lower than VBM. It is considered that such a K-poor condition is not allowed for unstable host materials. With reference to a previous study[13] where $Li_7TaO_6$ was obtained as a secondary phase during the synthesis, an additional stability condition was conducted. The stability condition of $KLi_6TaO_6$, not allowed by the additional $Li_7TaO_6$ condition ($7\mu_{Li} + \mu_{Ta} + 6\mu_O \leq \Delta H_f(Li_7TaO_6)$), is in agreement with the region where the host material is not stable because of the negative $E_f$ of $\square_K^-$. Figure 3d shows the results for condition *III*, which indicate K-rich, Li-rich, and Ta-poor conditions. Under this condition, $E_f$ of $Sn_{Ta}^-$ was significantly low, exhibiting negative values for the entire $E_F$. $E_F^{eq}$ was lower than that of VBM; therefore, the host material was also unstable. These results indicate that such a Ta-poor condition is not allowed.

Figure 3e shows the results for condition *IV*, which indicates K-rich, Li-intermediate, and Ta-intermediate conditions. The "intermediate" condition was assumed to be at the center of the rich and poor conditions. At $E_F^{eq}$, the lowest $E_f$ values for the positively and negatively charged defects were $Li_i^+$ and $Sn_{Ta}^-$, respectively. $E_f$ values of other defects were higher than those of $Li_i^+$ and $Sn_{Ta}^-$. Therefore, this condition corresponds to Li-excess conditions achieved by doping.

$E_f$ analyses for each $\mu$ condition were collected and are presented as scatter plots in Figure 3a. A successful achievement of Li-excess condition is considered the situation where the $c(Li_i^+)$ is more than twice as high as $c(\square_{Li}^-)$. Here, twice the concentration is selected as a sufficiently high value so that $Li_i^+$ can survive recombination with $\square_{Li}^-$. The results indicate that the formation of defects strongly depends on $\mu$ conditions, and one can tune the major defects by optimizing the synthesis condition.[31] For the stability of the host material, K-poor condition is prohibited, so that other K-excluding secondary phases, such as $Li_7TaO_6$, can be easily formed. Since controlling the loss of K is difficult,[13,25] maintaining a K-rich condition is important.

The Li-excess condition can be achieved using two methods. One method is doping (presented as a blue circle), the method desired. However, this is not achieved under Ta-rich condition. $\square_{Li}^-$ is easily formed when both K and Ta are rich (Li is poor). In a previous report,[13] Rietveld analysis exhibited that significant $\square_{Li}^-$ was indicated and sufficient amount of Sn was not substituted for Ta site. Probably, this indicated the formation of Li-poor and Ta-rich conditions. Therefore, maintaining K-rich and Li-rich conditions are required for Ta-poor condition. However, under extremely Ta-poor condition, the host material is not stable; therefore, optimization of a moderate level is required. Another method for achieving Li-excess condition is the formation of $Li_i^+$ without a dopant (presented as a green + symbol) by formation of $\square_K^-$. However, this is not preferred because a high $E_m$ is expected owing to the trapping of Li atom at the empty $\square_K^-$ site. More detailed results will be discussed later.

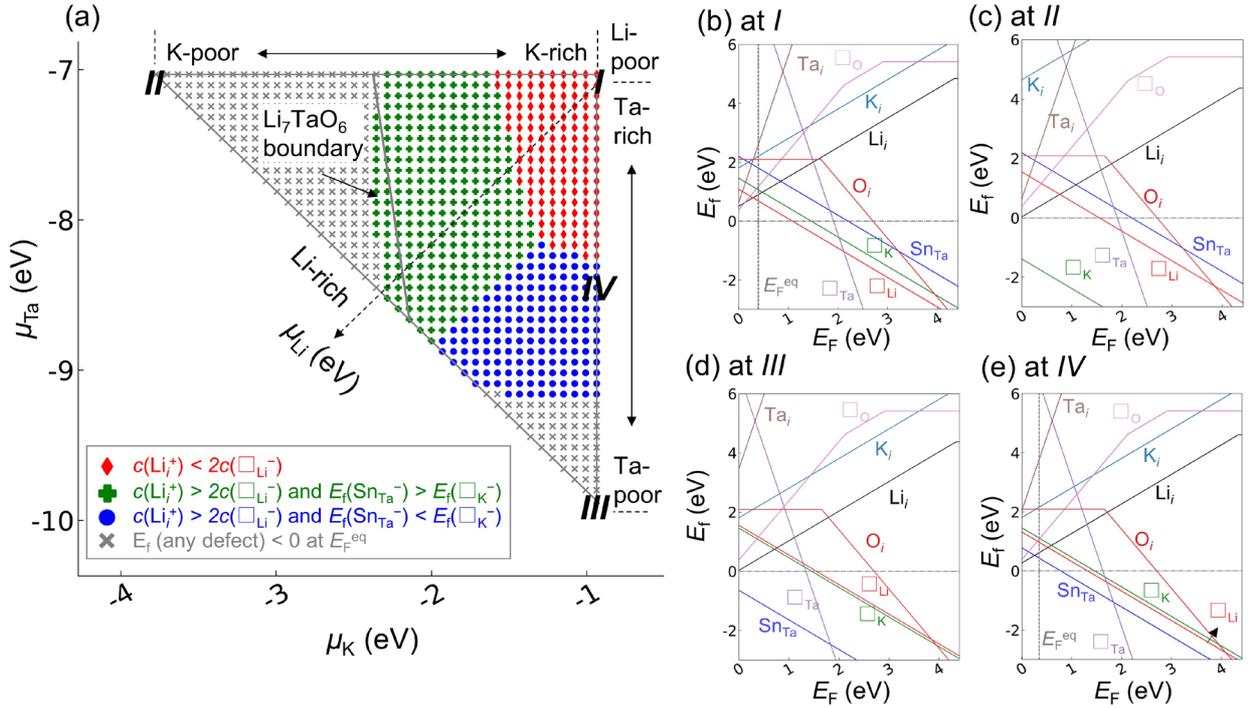

Figure 3. (a) The stability region (presented as a large triangle) of $KLi_6TaO_6$ in $\mu_K$ and $\mu_{Ta}$ at a temperature of 1073 K with limits imposed by binary oxides. Colored symbols inside the triangle indicate the condition relevant to the concentration and $E_f$ of point defects as variables of $\mu$. The red diamond symbol indicates $\mu$ conditions where the concentration of $Li_i^+$ is not twice as high as that of $\square_{Li}^-$. Green + and blue circle symbols indicate the chemical potential condition where the concentration of $Li_i^+$ is more than twice as high as that of $\square_{Li}^-$, but with the different lowest $E_f$ of negatively charged defects. The gray × symbol indicates $\mu$ conditions where $E_f$ of any defect is negative. It is assumed that $KLi_6TaO_6$ is not stable because the concentration of defects is larger than that of sites. For reference, another secondary phase such as $Li_7TaO_6$ boundary is investigated by the limit condition of $7\mu_{Li} + \mu_{Ta} + 6\mu_O \leq \Delta H_f(Li_7TaO_6)$. This reduces the allowed stability region of $KLi_6TaO_6$ from a large triangle to a smaller polygon. (b)–(e) $E_f$ of native defects and a dopant of $KLi_6TaO_6$ as a variable of $E_F$. $\mu$ is located for (b) at *I*, (c) at *II*, (d) at *III*, and (e) at *IV* in Figure 3a. The calculations were obtained by GGA parameterized in the PBE form. The gradient indicates the charge of a defect. The lowest $E_f$ obtained among various charges is displayed exclusively for each defect. $E_F^{eq}$ is an equilibrium Fermi level that is calculated by the charge neutrality condition.

Although it is assumed that the Li-excess condition is achieved by doping, the interaction between a dopant and $Li_i^+$ can decrease $\sigma_{Li}$. Figure 4 shows the $E_m$ values of Sn-doped $KLi_6TaO_6$ (in the $K_8Li_{49}SnTa_7O_{48}$ supercell) at various distances between $Li_i^+$ and $Sn_{Ta}^-$. Because of the interaction between oppositely charged defects, the energy where $Li_i^+$ is located close to $Sn_{Ta}^-$ (case *II*, and also shown in Supplementary Figure S2a) is lower than the energy where $Li_i^+$ is located far from $Sn_{Ta}^-$ by 0.29 eV. Hereinafter, this energy difference is referred to as $\Delta E_{far\text{-}close}$. The $\Delta E_{far\text{-}close}$, ascribed to the interaction between oppositely charged defects, increases $E_m$ to 0.33 eV. However, when the initial and final states used for the NEB calculations have the same interatomic distances between $Li_i^+$ and $Sn_{Ta}^-$, the $E_m$ value remains low (0.08–0.10 eV). The values are almost the same as those obtained without the dopant, regardless of whether the interatomic distances are short (case *I* and also shown in Supplementary Figure S2a) or long (cases *III–VI*). This implies that despite the existence of the dopant, Li atoms can migrate with a low $E_m$ when the interatomic distances between the defects are maintained after the migration of Li atoms. However, more energy is required when the Li atom at the interstitial site near the dopant escapes to other interstitial sites located far from the dopant (case *II*, and also shown in Supplementary Figure S2b). In fact, under a 2.1% Li-excess condition in Sn-doped $KLi_6TaO_6$ using $K_8Li_{49}SnTa_7O_{48}$, FPMD calculations estimate a $\sigma_{Li}$ of $8 \times 10^{-4}$ S/cm at 300 K, which is lower than that obtained without the dopant. The $E_a$ is 0.23 eV, which is between the lower $E_m$ of 0.08 eV and larger $E_m$ of 0.33 eV.

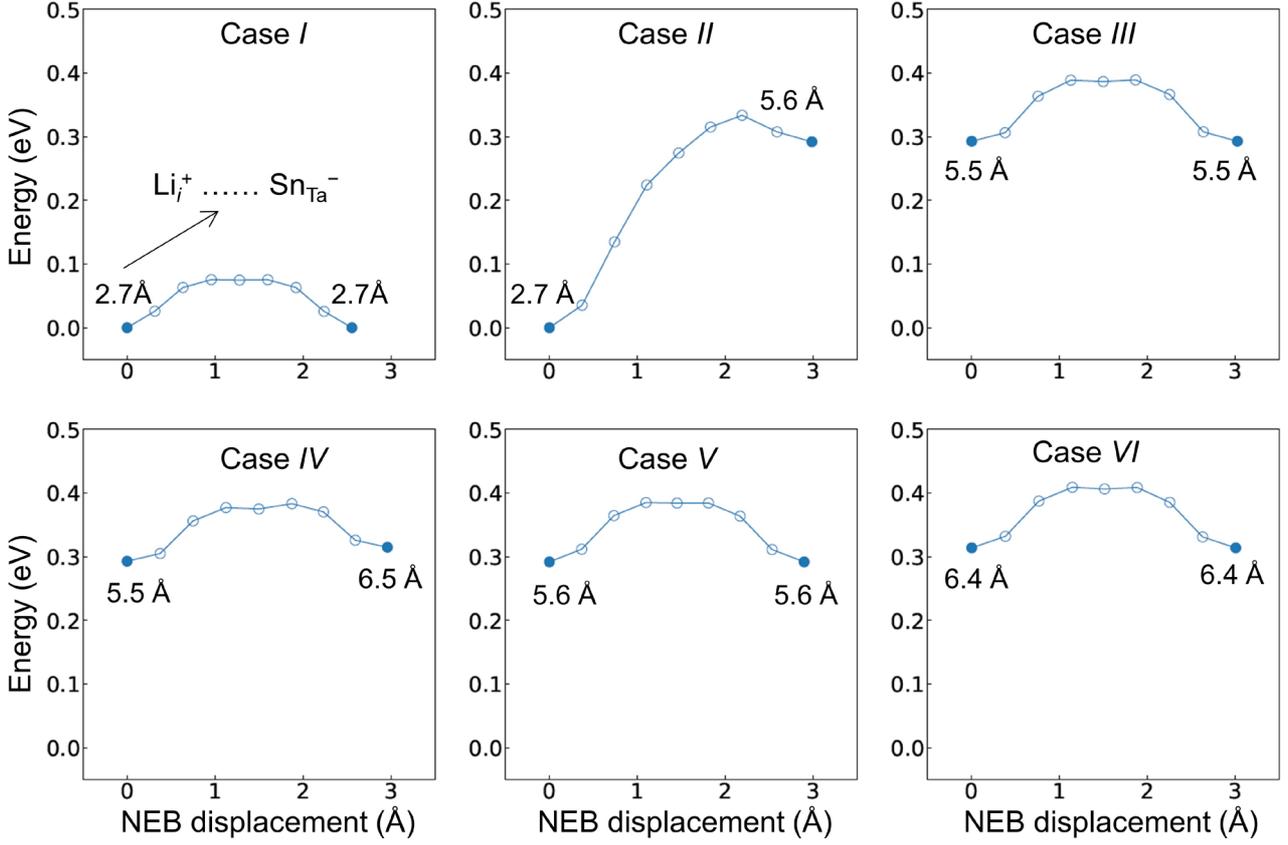

Figure 4. NEB profiles of $Li_i^+$ with different distances from $Sn_{Ta}^-$ in Sn-doped $KLi_6TaO_6$ (using a $2 \times 2 \times 2$ primitive cell with the formula $K_8Li_{49}SnTa_7O_{48}$). The numerical values written near the initial or final states indicate the distance between $Sn_{Ta}^-$ and $Li_i^+$. The energies of the states are shifted with respect to the most stable state where the distance between $Sn_{Ta}^-$ and $Li_i^+$ is approximately 2.7 Å. The closed and open circles indicate initial and final states where the atomic coordinates were fixed and optimized during the NEB calculations, respectively.

In addition, the interaction between $Li_i^+$ and $\square_K^-$ was also investigated when $Li_i^+$ is formed by $\square_K^-$ without doping. Supplementary Figure 3a shows NEB profiles of $Li_i^+$ with different distances from $\square_K^-$. The trend for the change in $E_m$ values is similar to that for the Sn-doped case: both initial and final states have long interatomic distances between the two oppositely charged defects, and $E_m$ values are small. However, an additional case exists here. Li atoms can be trapped at the $\square_K^-$ site when the trajectory for a moving Li atom is close to the empty $\square_K^-$ site. As shown in Supplementary Figure 3b, the Li atom moving *via* the kick-out mechanism escapes from the migration path shown in Figure 2a. The kicked-out Li atom is trapped at the $\square_K^-$ site, and the system energy is lowered. This increases the $E_m$ value significantly to 0.44 eV and indicates that the formation of $Li_i^+$ by doping is preferred to that by $\square_K^-$, to achieve a lower $E_m$ value.

Therefore, for material exploration of Li-ion conductors of the structure $A$Li$_6$XO$_6$ by complementing the weaknesses of KLi$_6$TaO$_6$, we need to consider the following: a better material requires a broad $\mu$ condition to possess a low $E_f$ of the dopant when compared with that of $\square_{Li}^-$, and a small $\Delta E_{far-close}$ relevant to interactions between the dopant and Li$_i^+$. The conditions for $A$Li$_6$XO$_6$ are discussed in Section 3.3.

**3.2. Primary screening results and suggestion of Li-ion conductors**

As described in Section 2.2, 58 combinations of various cations in $A$Li$_6$XO$_6$ were prepared with the same crystal structure. Material screening was performed based on three screening conditions: $\Delta H_f(A\text{Li}_6XO_6) < \Sigma\{\Delta H_f(\text{binary oxides})\}$, $w_{ph}^2 \geq 0$, and $E_m < 0.15$ eV. These are the primary conditions for achieving stable Li-ion conductors. The first condition indicates that the screened $A$Li$_6$XO$_6$ is more stable than the corresponding binary oxides. Here, as the binary oxides, the most stable structures reported in Materials Project Database (MPD)[47,48] were selected considering the valences of $A$ and $X$. For example, For KLi$_6$TaO$_6$, K$_2$O, Li$_2$O, and Ta$_2$O$_5$ were selected. The second condition indicates that the screened $A$Li$_6$XO$_6$ is dynamically stable without any imaginary phonon frequency. The third condition indicates that the screened $A$Li$_6$XO$_6$ is expected to have high $\sigma_{Li}$ under Li-excess conditions. Here, the value of 0.15 eV is approximately twice as high as that of KLi$_6$TaO$_6$ which has a sufficiently low $E_m$ of 0.08 eV.

Among them, the combination of trivalent cations that occupied both the $A$ (3$b$ site) and $X$ (3$a$ site) sites was excluded by the screening condition of $\Delta H_f(A\text{Li}_6XO_6) < \Sigma\{\Delta H_f(\text{binary oxides})\}$. The combinations of divalent and tetravalent cations that occupied the $A$ and $X$ sites, respectively, were excluded by the screening condition of $w_{ph}^2 \geq 0$. The oxides with Na or Ag occupying the $A$ site were also excluded under the three screening conditions. $E_m$ values of 34 $A$Li$_6$XO$_6$ types are summarized in Supplementary Table S1. These oxides were more stable than the mixture of corresponding oxides because they satisfy the screening condition of $\Delta H_f(A\text{Li}_6XO_6) < \Sigma\{\Delta H_f(\text{binary oxides})\}$. In addition, their $E_m$ could exhibit reasonable values obtained from the migration path as shown in Figure 2a under the Li-excess condition. Supplementary Figure S4 shows the correlation of $E_m$ with ionic radii[49,50] of $A$ ($R_A$), $X$ ($R_X$), and the difference between $A$ and $X$ ($R_A - R_X$), with linear correlation coefficients ($r$) of −0.62, 0.53, and −0.78, respectively. A larger $R_A$ and a smaller $R_X$ give a smaller $E_m$.

Twelve oxides with K or Rb atoms at the *A* site satisfied all three screening conditions. They are more stable than the mixture of corresponding binary oxides. Among them, $KLi_6TaO_6$[13,25] and $KLi_6BiO_6$[51] have been synthesized, and the order of magnitude of $\sigma_{Li}$ at 300 K under Li-excess conditions[13,15,16] was estimated to be $10^{-3}$–$10^{-2}$ S/cm by FPMD. In addition, according to the phase diagram in MPD,[47,48] these two oxides are confirmed to be a convex–hull state when oxygen chemical potential corresponds to the value at the temperature of 1073 K. Therefore, they are thermodynamically more stable than other ternary or quaternary oxides that have been already calculated. The dynamical stabilities of the 12 oxides are confirmed by $w_{ph}^2 \geq 0$, as shown in Supplementary Figure S5. The $E_m$ values (< 0.15 eV) are summarized in Table 1. The NEB profiles are shown in Supplementary Figure S6. The $E_a$ and predicted $\sigma_{Li}$ values at 300 K obtained from the FPMD calculations are summarized in Table 1. The linear relationship between MSD and simulation time obtained by FPMD calculations is shown in Supplementary Figure S7.

All 12 oxides in the list are insulators. The theoretical lattice parameters calculated by the GGA method in PBE functional[24] are in agreement with the previously reported experimental and theoretical values, within a difference of only 1.3%. In these combinations, when K is exchanged by Rb, the volume per atom increases by 1.6–2.3%. However, for $KLi_6XO_6$ and $RbLi_6XO_6$ with the same cation $X$ under Li-excess conditions, $E_m$ is similar within a difference of 0.02 eV. The $E_g$ values are similar within a difference of 0.1 eV.

The 12 oxides are considered to be promising Li-ion conducting oxides with low $E_m$ and $E_a$ values. Their $E_m$ values range from 0.04 eV to 0.13 eV. Among $KLi_6XO_6$ structures, the smaller the volume or lattice parameter is, the smaller the $E_m$ value tends to be. The $E_a$ values range from 0.09 eV to 0.19 eV. The $E_m$ and $E_a$ values agreed with each other, despite the slightly larger value of $E_a$. Considering a similar $E_m$ between $KLi_6XO_6$ and $RbLi_6XO_6$ with the same cation $X$, the larger deviation of $E_a$ might be ascribed to the statistical deviations in the FPMD calculations. Based on the similar values of $E_m$ and $E_a$, and the consistency between the NEB and FPMD trajectories of $KLi_6TaO_6$ confirmed in a previous study,[13] we consider that the main migration path for the Li atoms under the Li-excess condition is in agreement with the suggested path in Figure 2a. The estimated $\sigma_{Li}$ at 300 K under Li-excess conditions and the minimum value obtained from the error range (one standard deviation obtained from three different FPMD trials) are in the order of magnitude of $10^{-3}$–$10^{-2}$ S/cm. The estimated $\sigma_{Li}$ at 300 K ($4\times10^{-3}$ S/cm) of $KLi_6BiO_6$ with additional $Li_i$ is consistent with the value ($6\times10^{-3}$ S/cm) reported in a previous FPMD study for $KLi_{6.5}Sn_{0.5}BiO_6$ (also under Li-excess conditions).[15,16] It is noteworthy that these oxides also require Li-excess conditions to achieve such a low $E_m$. Under stoichiometric conditions, the atomic coordinates with an intentionally prepared pair of $Li_i^+$ and $\square_{Li}^-$ were not dynamically stable, and the $Li_i^+$ and $\square_{Li}^-$ pair recombined soon.

Note that (K or Rb)Li$_6$VO$_6$ and (K or Rb)Li$_6$AsO$_6$ have the most stable intermediate states, with energies lower than those of the initial and final states, as shown in Supplementary Figure S6. The characteristics of the NEB profiles of these oxides are different from those of the other 8 oxides, which have the initial or final states at lower energies relative to those of the intermediate states. However, for (K or Rb)Li$_6$VO$_6$ and (K or Rb)Li$_6$AsO$_6$, the most stable intermediate state with the lowest energy has two Li atoms at two interstitial sites and one $\square_{Li}^-$, which is very similar to state *iii* of KLi$_6$TaO$_6$ shown in Figure 2b. This means that (K or Rb)Li$_6$VO$_6$ and (K or Rb)Li$_6$AsO$_6$ thermodynamically prefer forming two $Li_i^+$ ions and one $\square_{Li}^-$, instead of forming one $Li_i^+$ ion at the 9$d$ interstitial site under Li-excess conditions.

Table 1. Main properties of the 12 suggested $A$Li$_6$XO$_6$ structures.

| Material | Lattice parameter (Å)[a] | | $E_g$ (eV) | | $\Delta H_f$ (eV/f.u.)[b] | $E_m$ (eV) | $E_a$ (eV)[c,h] | Estimated $\sigma_{Li}$ at 300 K (mS/cm)[b,h] |
|---|---|---|---|---|---|---|---|---|
| | $a$ and $b$ | $c$ | by PBE | by HSE06 | | | | |
| KLi$_6$VO$_6$ | 8.106 | 7.108 | 1.81 | 3.24 | −28.30 (−1.86) | 0.04 | 0.11 (0.10,0.11) | 62 (58,67) |
| RbLi$_6$VO$_6$ | 8.155 | 7.182 | 1.80 | 3.22 | −28.02 (−1.77) | 0.05 | 0.16 (0.16,0.17) | 18 (15,22) |
| KLi$_6$NbO$_6$ | 8.274 | 7.263 | 3.74 | 5.26 | −30.55 (−2.51) | 0.08 | 0.11 (0.11,0.12) | 44 (41,45) |
| RbLi$_6$NbO$_6$ | 8.321 | 7.318 | 3.70 | 5.21 | −30.31 (−2.46) | 0.07 | 0.14 (0.11,0.19) | 22 (5,56) |
| KLi$_6$TaO$_6$ | 8.283 (8.226,[d] 8.224,[e] 8.317[f]) | 7.265 (7.212,[d] 7.290[f]) | 4.41 (4.41[f]) | 5.97 | −31.23 (−2.84) | 0.09 | 0.11 (0.11,0.12) | 36 (24,50) |
| RbLi$_6$TaO$_6$ | 8.326 | 7.320 | 4.37 | 5.91 | −30.99 (−2.79) | 0.08 | 0.14 (0.13,0.15) | 22 (13,33) |
| KLi$_6$AsO$_6$ | 8.143 | 7.009 | 2.61 | 4.34 | −25.07 (−2.95) | 0.06 | 0.13 (0.13,0.13) | 45 (40,50) |
| RbLi$_6$AsO$_6$ | 8.176 | 7.109 | 2.52 | 4.24 | −24.75 (−2.82) | 0.07 | 0.18 (0.16,0.19) | 17 (15,21) |
| KLi$_6$SbO$_6$ | 8.340 | 7.225 | 3.06 | 4.74 | −26.39 (−3.62) | 0.08 | 0.09 (0.09,0.10) | 77 (69,86) |
| RbLi$_6$SbO$_6$ | 8.382 | 7.287 | 3.06 | 4.73 | −26.15 (−3.58) | 0.06 | 0.17 (0.15,0.20) | 9 (3,19) |
| KLi$_6$BiO$_6$ | 8.526 (8.416,[g] 8.569[f]) | 7.383 (7.294,[g] 7.410[f]) | 1.45 (1.43[f]) | 2.79 | −24.11 (−2.70) | 0.13 | 0.19 (0.19,0.20) | 4 (3,5) |
| RbLi$_6$BiO$_6$ | 8.573 | 7.427 | 1.43 | 2.77 | −23.94 (−2.72) | 0.12 | 0.10 (0.09,0.11) | 47 (32,65) |

[a] From a conventional cell with a hexagonal structure ($A_3$Li$_{18}X_3$O$_{18}$). The lattice angles, $\alpha = \beta = 90°$, and $\gamma = 120°$.
[b] The abbreviation f.u. stands for formula unit. The values in parentheses are $\Delta H_f$ ($A$Li$_6$XO$_6$)–Σ{$\Delta H_f$ (binary oxides)}. Binary oxides used were K$_2$O, Li$_2$O, Rb$_2$O, V$_2$O$_5$, Nb$_2$O$_5$, Ta$_2$O$_5$, As$_2$O$_5$, and Sb$_2$O$_5$. For Bi, the coexistence of BiO$_2$ + 0.5O$_2$ was used, which is more stable than Bi$_2$O$_5$.

[c] Li-excess condition with a concentration of +2.1% (49/48) was used.
[d] From Ref. [25] (experiment).   [e] From Ref. [13] (Experiment).   [f] From MPD (theory)[47,48].
[g] From Ref. [51] (experiment).
[h] Values in parentheses are error ranges. The lower and upper values were obtained using ± one standard deviation, calculated from three different FPMD trials.

### 3.3. Doping effect on the suggested Li-ion conductors

As mentioned in Section 3.1, KLi$_6$TaO$_6$ has two weaknesses that can hinder the achievement of the high $\sigma_{Li}$ estimated by the theoretical calculations: First, forming a Li-excess condition by doping is limited because of lower $E_f$ of other negatively charged defect such as $\square_{Li}^-$, and second, the negatively charged extrinsic dopant interacts with $Li_i^+$ and increases $E_m$. Therefore, it is worthwhile to investigate the doping effects on the $E_f$ and $E_m$ values of the suggested Li-ion conducting oxides by inserting various dopants.

Table 2 shows the $E_f$ of a complex point defect of $(M_X + Li_i)^0$ with various types of $M$ under $M$-rich and Li-rich condition. KLi$_6$XO$_6$ and RbLi$_6$XO$_6$ have similar $E_f$ values for the same dopant. In most cases, the lowest $E_f$ is obtained by Sn substitution among the various types of $M$. The $E_f$ value of Pb substitution is higher than that of Sn substitution by 0.0–0.2 eV, except for (K or Rb)Li$_6$BiO$_6$. For (K or Rb)Li$_6$BiO$_6$, Pb substitution exhibits a slightly lower $E_f$ than Sn substitution by approximately 0.1 eV. (K or Rb)Li$_6$XO$_6$, with a relatively small radius of $X$ = (V or As), has $E_f$ values of Ti or Ge substitution that are higher than that of Sn substitution by 0.0–0.2 eV. The $E_f$ values of Zr, Hf, and Ce substitution are much higher than that of Sn substitution by >0.2 eV. The selection of Sn as a dopant is reasonable among the various dopants considered. Therefore, it is necessary to investigate the $\mu$ condition where negatively charged Sn$_X^-$ can be stabilized more than than $\square_{Li}^-$.

Table 2. $E_f$ of the complex point defects of $(M_X + Li_i)^0$ with various types of $M$ of the suggested $A$Li$_6$XO$_6$ structures. The values in bold are the lowest ones among the various types of $M$. $\mu$ for $X$ and Li are set at $X$-rich and Li-rich conditions.

| Combination of $A$ and $X$ | Dopant, $M$ | | | | | | |
| --- | --- | --- | --- | --- | --- | --- | --- |
| | Ge | Sn | Pb | Ti | Zr | Hf | Ce |
| K, V | 1.11 | **1.06** | 1.25 | 1.16 | 1.41 | 1.29 | 2.37 |
| Rb, V | 1.08 | **0.95** | 1.10 | 1.12 | 1.31 | 1.20 | 2.24 |
| K, Nb | 1.84 | **1.61** | 1.68 | 1.83 | 1.93 | 1.85 | 2.72 |
| Rb, Nb | 1.86 | **1.56** | 1.59 | 1.84 | 1.90 | 1.81 | 2.65 |

| | | | | | | |
|---|---|---|---|---|---|---|
| K, Ta | 2.17 | **1.95** | 2.01 | 2.17 | 2.27 | 2.18 | 3.05 |
| Rb, Ta | 2.20 | **1.90** | 1.93 | 2.18 | 2.24 | 2.15 | 3.00 |
| K, As | 2.18 | **2.15** | 2.35 | 2.20 | 2.49 | 2.37 | 3.50 |
| Rb, As | 2.02 | **1.96** | 2.11 | 2.07 | 2.30 | 2.20 | 3.27 |
| K, Sb | 2.96 | **2.69** | 2.74 | 2.92 | 3.00 | 2.92 | 3.77 |
| Rb, Sb | 2.97 | **2.64** | 2.64 | 2.92 | 2.96 | 2.88 | 3.68 |
| K, Bi | 1.61 | 1.18 | **1.09** | 1.53 | 1.47 | 1.41 | 2.03 |
| Rb, Bi | 1.72 | 1.21 | **1.08** | 1.62 | 1.51 | 1.46 | 2.02 |

Figure 5 shows the stability condition of the suggested KLi$_6$XO$_6$ structures at a temperature of 1073 K and the major defect information obtained from the concentration and $E_f$ of point defects as variables of $\mu$. Note that, herein, only the results of KLi$_6$XO$_6$ are shown because the $E_f$ values of KLi$_6$XO$_6$ and RbLi$_6$XO$_6$ with the same $X$ are almost identical (see Supplementary Figure S8). Similar to the characteristics of KLi$_6$TaO$_6$ as shown in Figure 3a, these oxides also have limited $\mu$ conditions to form Li-excess conditions by doping. Under $X$-rich conditions, $E_f$ of $M_X^-$ is higher than that of other negatively charged defects such as $\square_{Li}^-$ and $\square_K^-$. Therefore, $M$ is not substituted easily for $X$. The K-rich condition has to be maintained because K is lost during the synthesis.[13,25] Under K-rich and $X$-poor conditions at a moderate level, the formation of Li-excess conditions is possible by doping, except for KLi$_6$VO$_6$. Among the six oxides, KLi$_6$BiO$_6$ has a wide range of $\mu$ conditions to form Li-excess conditions by doping. As an example, under one allowed $\mu$ condition (presented as a symbol * in Figure 5), the change in $E_f$ as a variable of $E_F$ of the suggested KLi$_6$XO$_6$ is plotted in Supplementary Figure S9.

KLi$_6$VO$_6$ showed different characteristics from the other oxides as this oxide has low $E_f$ values for other defects, such as $\square_O^{2+}$ and $V_i^{5+}$. Accordingly, the allowed $\mu$ conditions are significantly constrained. Any rich condition is not allowed to form Li-excess condition by doping; therefore, optimization of $\mu$ is more challenging than the other suggested KLi$_6$XO$_6$.

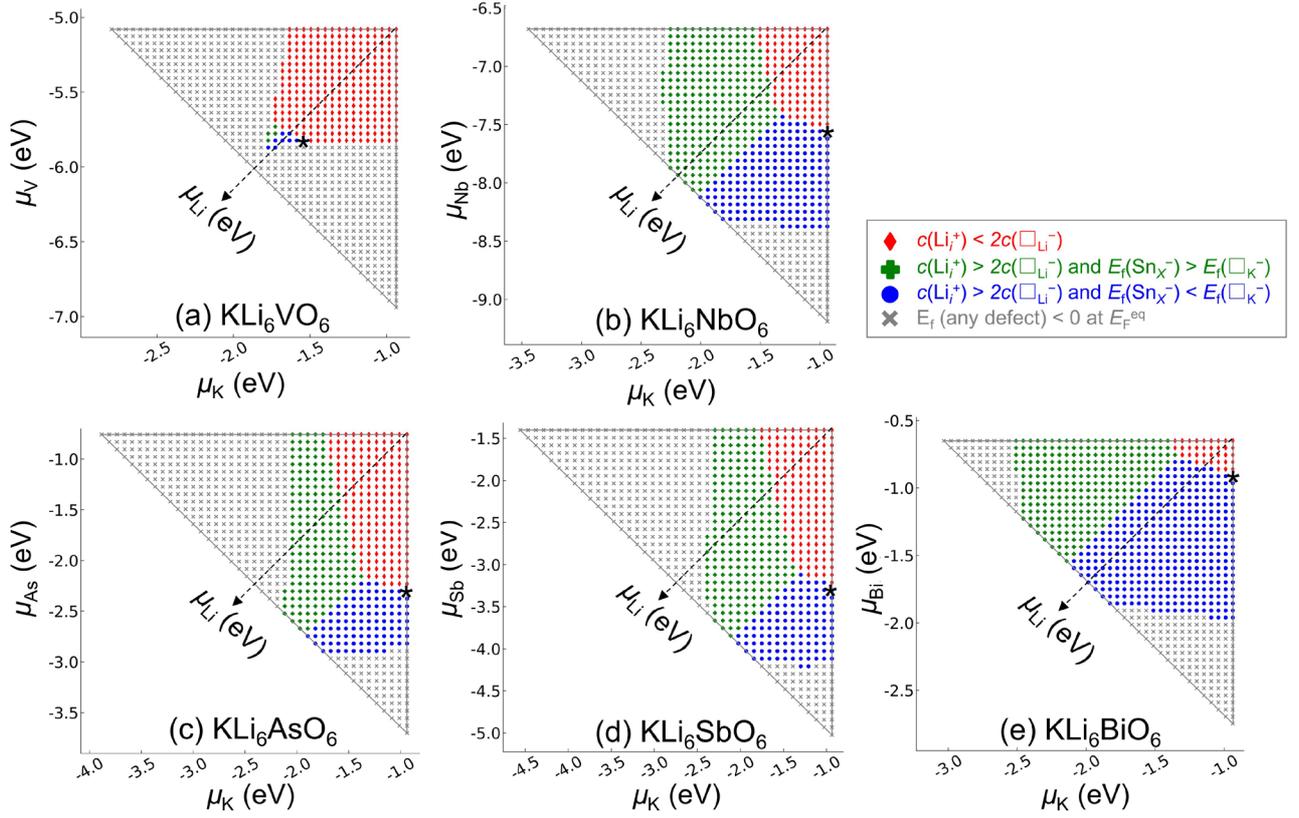

Figure 5. The stability region (triangle) of the suggested $KLi_6XO_6$ structures, (except for $KLi_6TaO_6$, see Figure 3a) (a) $KLi_6VO_6$, (b) $KLi_6NbO_6$, (c) $KLi_6AsO_6$, (d) $KLi_6SbO_6$, and (e) $KLi_6BiO_6$ in $\mu_K$ and $\mu_X$ at a temperature of 1073 K with limits imposed by binary oxides. Colored symbols inside the triangle indicate the condition relevant to the concentration and $E_f$ of point defects as variables of $\mu$. The calculations for $E_f$ were obtained by GGA parameterized in the PBE form. The detailed information for each symbol can be referred to in Figure 3a. The symbol * indicates a $\mu$ condition for an available example of the Li-excess condition by doping for a plot of $E_f$ as a variable of $E_F$ (see Supplementary Figure S9).

The doping effect on the $E_m$ values of the suggested $ALi_6XO_6$ (using $A_8Li_{49}X_7MO_{48}$ supercell) was analyzed by assuming that a Li-excess condition was achieved. As mentioned in the descriptions in Figure 4, because of the interaction between negatively charged $M_X^-$ and $Li_i^+$, $Li_i^+$ experiences a potential energy of $\Delta E_{far-close}$, moving far away from $M_X^-$, which results in an increase in $E_m$. Table 3 shows $\Delta E_{far-close}$ and $E_m$. Compared with the dependence of $E_f$ on the type of $M$, that of $\Delta E_{far-close}$ is much smaller (approximately 0.1 eV). The $E_m$ values range from 0.16 to 0.38 eV, which are much higher than the values obtained without the dopants (0.04–0.13 eV as summarized in Table 1). In most cases, the difference of $E_m$ and $\Delta E_{far-close}$ is only <0.07 eV (except for Ce-doped $KLi_6BiO_6$ with a difference of 0.1 eV), so that the contribution of $\Delta E_{far-close}$ is significant for increasing $E_m$.

Table 3. $\Delta E_{\text{far-close}}$ values, corresponding to case *II* shown in Figure 4, of the suggested $A\text{Li}_6X\text{O}_6$ structures. The values in parentheses are $E_m$ values.

| Combination of $A$ and $X$ | Dopant, $M$ | | | | | | |
|---|---|---|---|---|---|---|---|
| | Ge | Sn | Pb | Ti | Zr | Hf | Ce |
| K, V | 0.20 (0.21) | 0.20 (0.21) | 0.24 (0.26) | 0.22 (0.22) | 0.19 (0.20) | 0.18 (0.19) | 0.22 (0.24) |
| Rb, V | 0.20 (0.20) | 0.19 (0.19) | 0.22 (0.23) | 0.20 (0.20) | 0.17 (0.17) | 0.16 (0.16) | 0.20 (0.21) |
| K, Nb | 0.31 (0.33) | 0.27 (0.31) | 0.30 (0.35) | 0.32 (0.35) | 0.24 (0.29) | 0.23 (0.28) | 0.25 (0.32) |
| Rb, Nb | 0.31 (0.32) | 0.26 (0.29) | 0.29 (0.33) | 0.34 (0.34) | 0.23 (0.26) | 0.22 (0.25) | 0.23 (0.29) |
| K, Ta | 0.34 (0.36) | 0.29 (0.33) | 0.33 (0.38) | 0.36 (0.38) | 0.27 (0.31) | 0.26 (0.30) | 0.27 (0.34) |
| Rb, Ta | 0.34 (0.35) | 0.29 (0.31) | 0.32 (0.35) | 0.35 (0.36) | 0.26 (0.29) | 0.25 (0.28) | 0.25 (0.31) |
| K, As | 0.29 (0.29) | 0.28 (0.29) | 0.33 (0.34) | 0.32 (0.32) | 0.28 (0.29) | 0.26 (0.27) | 0.30 (0.32) |
| Rb, As | 0.17 (0.17) | 0.24 (0.25) | 0.29 (0.29) | 0.29 (0.29) | 0.23 (0.24) | 0.22 (0.22) | 0.25 (0.27) |
| K, Sb | 0.34 (0.35) | 0.28 (0.32) | 0.32 (0.36) | 0.36 (0.37) | 0.27 (0.30) | 0.26 (0.29) | 0.27 (0.33) |
| Rb, Sb | 0.35 (0.34) | 0.27 (0.29) | 0.30 (0.33) | 0.35 (0.35) | 0.25 (0.28) | 0.24 (0.27) | 0.24 (0.30) |
| K, Bi | 0.30 (0.33) | 0.22 (0.29) | 0.25 (0.32) | 0.32 (0.35) | 0.21 (0.27) | 0.20 (0.26) | 0.18 (0.29) |
| Rb, Bi | 0.31 (0.32) | 0.23 (0.27) | 0.25 (0.31) | 0.33 (0.35) | 0.20 (0.25) | 0.19 (0.24) | 0.17 (0.24) |

For the suggested oxides, the dependence of $\Delta E_{\text{far-close}}$ or $E_m$ on a characteristic property, in particular, a fourth ionization energy[52] of the dopants, was investigated and is shown in Supplementary Figure S10. For doped (K or Rb)$\text{Li}_6X\text{O}_6$ ($X$ = Ta, Nb, Sb, or Bi), $\Delta E_{\text{far-close}}$ or $E_m$ is strongly correlated with the fourth ionization energy of the tetravalent dopants, with absolute values of $r \geq 0.85$. This strong correlation could be ascribed to the depth of the electrostatic potential at the dopants. A common feature of these oxides is that they have the lowest energy on the potential energy surface for the initial and final states, with an excess Li atom occupying the $9d$ interstitial site (see Supplementary Figure S6). Among the dopants, the substitution by Hf, with the lowest fourth ionization energy, results in a lower $\Delta E_{\text{far-close}}$ or $E_m$ than that of other dopants. However, for doped (K or Rb)$\text{Li}_6$(V or As)$\text{O}_6$, $\Delta E_{\text{far-close}}$ or $E_m$ values do not have a strong correlation with the fourth ionization energy of the tetravalent dopants. These oxides had the lowest potential energy, not at the initial or final state, but at the most stable intermediate state, which has two Li atoms occupying the interstitial sites with remaining one Li vacancy site. $\Delta E_{\text{far-close}}$ is obtained from the energy difference of the initial and final states, but not the intermediate states; therefore, for these oxides, other factors, such as different potential energy surfaces, might affect the dependence of $\Delta E_{\text{far-close}}$ on the type of dopant.

KLi$_6$BiO$_6$ showed a wider range of $\mu$ conditions for achieving Li-excess conditions by doping than the other KLi$_6$XO$_6$. The $E_m$ of (K or Rb)Li$_6$BiO$_6$ is slightly lower than that of Sn-doped KLi$_6$TaO$_6$. The doped (K or Rb)Li$_6$VO$_6$ showed a lower $E_m$ than the other doped (K or Rb)Li$_6$XO$_6$. Moreover, the $E_m$ values of (K or Rb)Li$_6$VO$_6$ are lower than that of KLi$_6$TaO$_6$ by approximately 0.1 eV. We performed FPMD to estimate the $\sigma_{Li}$ at 300 K for KLi$_6$VO$_6$ and KLi$_6$BiO$_6$ under Li-excess conditions by substituting Sn for V or Bi. The properties of the Rb-containing oxides are considered similar to those of the corresponding K-containing oxides because of their similar $E_m$ values. To compare the results with those of Sn-doped KLi$_6$TaO$_6$,[13] the same temperature conditions were employed for FPMD. Supplementary Figure S11 shows the FPMD results for $\sigma_{Li}$. It shows a linear relationship between MSD and the simulation time. In the result, the Sn-doped KLi$_6$BiO$_6$ shows slightly lower $E_a$ (0.20 eV) and higher $\sigma_{Li}$ at 300 K ($2\times10^{-3}$ S/cm) than those of the Sn-doped KLi$_6$TaO$_6$ ($E_a$: 0.23 eV and $\sigma_{Li}$: $7\times10^{-4}$ S/cm). Sn-doped KLi$_6$VO$_6$ shows a lower $E_a$ (0.15 eV) and a much higher $\sigma_{Li}$ at 300 K ($2\times10^{-2}$ S/cm) than those of the other two materials. Note that $E_a$ lies between the $E_m$ values obtained with and without the interaction between the dopant and Li$_i^+$. The error range, obtained by one standard deviation from different FPMD trials, is wider than that obtained by Li-excess (K or Rb)Li$_6$XO$_6$ without a dopant. This is probably because the potential surface becomes more complex owing to the incorporated dopant. The lowest estimated $\sigma_{Li}$ at 300 K from the error range of Sn-doped KLi$_6$VO$_6$, KLi$_6$BiO$_6$, and KLi$_6$TaO$_6$ are $8\times10^{-3}$, $3\times10^{-4}$, and $3\times10^{-4}$ S/cm, respectively. Since low $E_m$ obtained using the NEB method for doped (K or Rb)Li$_6$XO$_6$ are similar to those of Sn-doped KLi$_6$BiO$_6$, the estimated $\sigma_{Li}$ at 300 K is expected to be higher than $10^{-4}$ S/cm. The FPMD result, the low $E_m$ obtained using the NEB method, and the $E_f$ results suggest that the suggested oxides are promising Li-ion conductors under Li-excess conditions.

## 4. Conclusion

In this study, the Li-ion migration mechanism and doping effect of KLi$_6$TaO$_6$ were analyzed in detail. The $E_m$ of Li$_i^+$ *via* the kick-out mechanism (0.08 eV) was much lower than that of $\square_{Li}^-$. This indicates that a Li-excess condition is essential for obtaining a low $E_m$. However, $E_f$ values revealed that the incorporation of Li$_i^+$ by doping (substituting tetravalent element for Ta) is possible under particular chemical potential conditions. If an adequate chemical potential condition is not achieved, the formation of negatively charged defects, such as $\square_{Li}^-$, is preferred over the dopant. $\square_{Li}^-$ recombines with $Li_i^+$ and prevent obtaining a low $E_m$. Although the doping was successful, the interaction of the dopant and $Li_i^+$ raised $E_m$ from 0.08 to 0.32 eV when their interatomic distances increased. These two weaknesses can explain why the experimentally measured $\sigma_{Li}$ was much lower than that estimated by the FPMD calculations in a previous study.[13]

We performed primary screening of 58 types of $A$Li$_6X$O$_6$ structures (valence sum of $A$ and $X$ is 6), which are isostructural with KLi$_6$TaO$_6$, on the basis of stability and low $E_m$. As a result, 12 types of (K or Rb)Li$_6X$O$_6$, where $X$ is pentavalent, have been suggested. Their $E_m$, with the migration paths connected in three dimensions, was significantly low, <0.15 eV. KLi$_6X$O$_6$ and RbLi$_6X$O$_6$ with the same type $X$ structure had properties similar to those of defect formation and Li-ion migration.

$E_f$ values suggested that the 12 oxides have particular chemical potential conditions to achieve Li-excess conditions by doping. The chemical potential conditions and relevant $E_f$ of various defects strongly depend on the synthesis conditions; therefore, the major defects can be tuned by optimizing the synthesis conditions. These 12 oxides also have the interaction between $Li_i^+$ and dopant, which increases $E_m$. Combinations of the suggested isostructural oxides and dopants are identified to reduce the interactions between $Li_i^+$ and dopant. FPMD calculations were performed for Sn-doped KLi$_6$VO$_6$ and KLi$_6$BiO$_6$ with a low and high $E_m$ among the suggested oxides, respectively. The results show that Sn-doped KLi$_6$VO$_6$ and KLi$_6$BiO$_6$ exhibited $\sigma_{Li}$ of $10^{-3}$–$10^{-2}$ S/cm at 300 K. These results suggested that the oxides under Li-excess conditions achieved by doping could be good Li-ion conductors.

## ■ AUTHOR INFORMATION

**Corresponding Author**

Joohwi Lee − Toyota Central R&D Laboratories, Inc., Nagakute, Aichi 480-1192, Japan; orcid.org/0000-0002-4263-471X; Email: j-lee@mosk.tytlabs.co.jp

**Authors**

Ryoji Asahi − Toyota Central R&D Laboratories, Inc., Nagakute, Aichi 480-1192, Japan; Present address : Nagoya University, Nagoya, Aichi 464-8603, Japan; orcid.org/0000-0002-2658-6260

## ■ ACKLOWLEDGEMENTS

The authors thank Dr. N. Ohba in TCRDL for fruitful discussion and Editage (www.editage.com) for English language editing.

**Supporting Information**

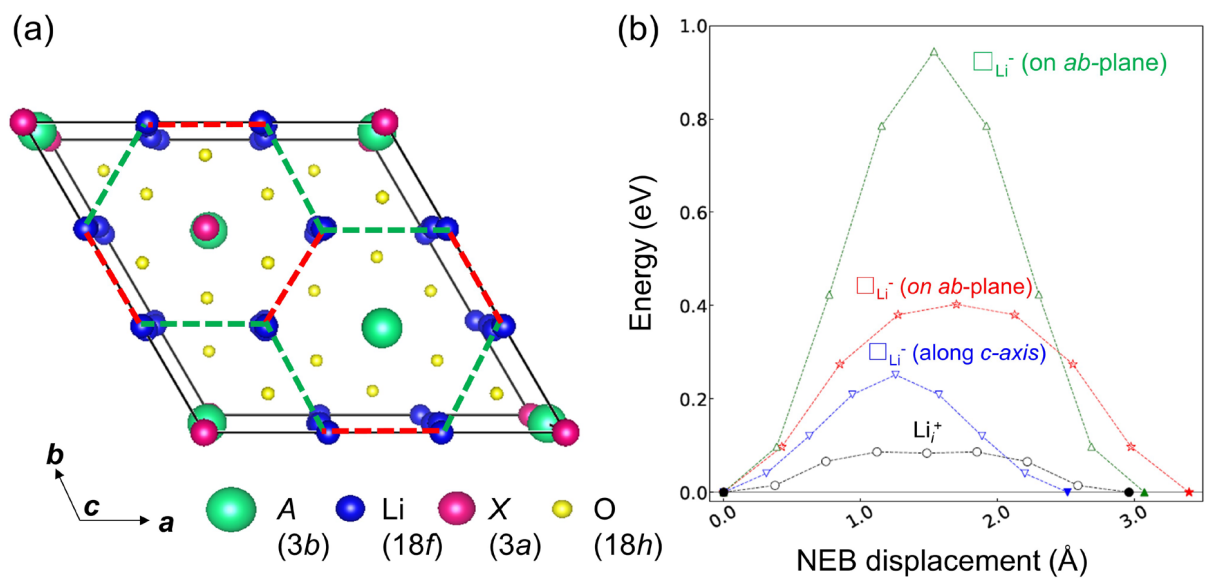

Figure S1. (a) Migration path for $\square_{Li}^{-}$ in the $A$Li$_6X$O$_6$ structure. Green and red dashed lines indicate two types of paths on the $ab$-plane. There is another migration path along the $c$-axis not shown in this figure. (b) Comparison of the NEB profiles of KLi$_6$TaO$_6$ for $\square_{Li}^{-}$ under Li-deficient conditions and Li$_i^{+}$ under Li-excess conditions. The closed and open symbols indicate initial and final states where the atomic coordinates were fixed and optimized during the NEB calculations, respectively.

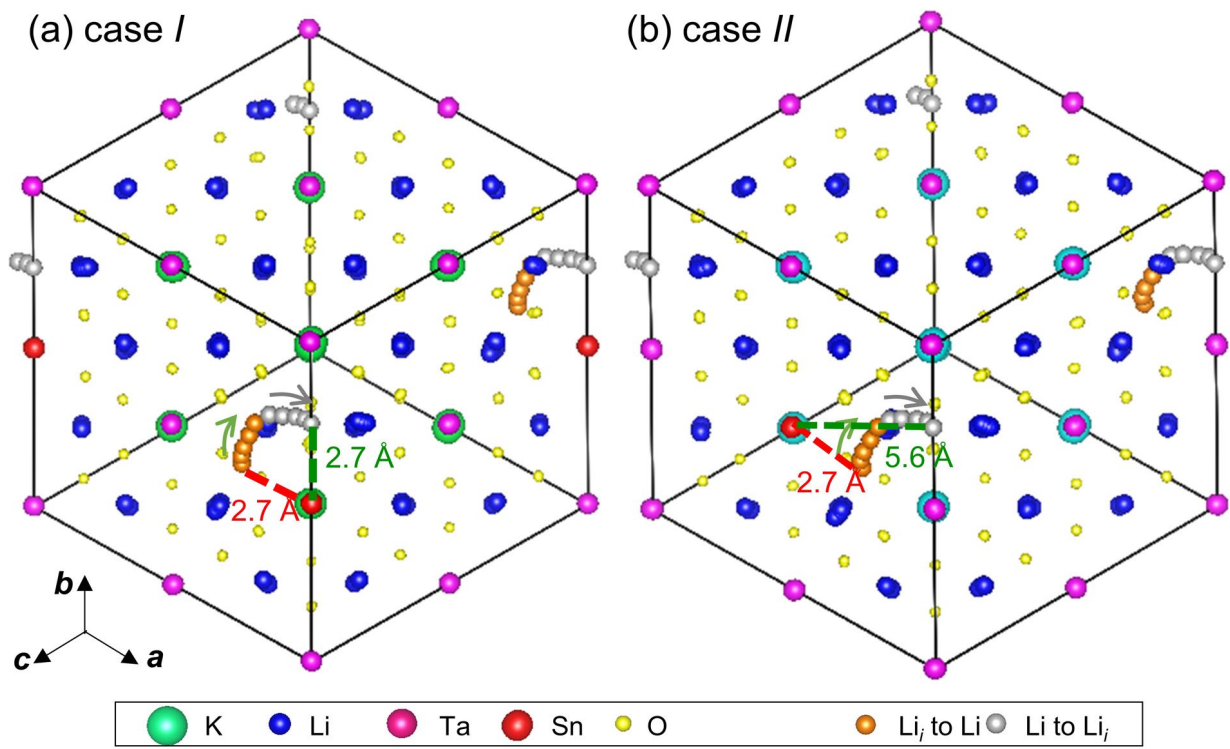

Figure S2. Trajectory of the moving atoms during the NEB calculations involving Sn-doped $KLi_6TaO_6$ under Li-excess conditions (using a 2 × 2 × 2 primitive cell with the formula $K_8Li_{49}SnTa_7O_{48}$) for cases *I* and *II* shown in Figure 4. Red and green dashed lines indicate the distance between $Sn_{Ta}^-$ and $Li_i^+$ in the initial and final states, respectively. In case *I*, the interatomic distance between $Sn_{Ta}^-$ and $Li_i^+$ in the initial and final states is the same (2.7 Å), whereas in case *II*, the interatomic distance between $Sn_{Ta}^-$ and $Li_i^+$ in the final state is double that observed in the initial state.

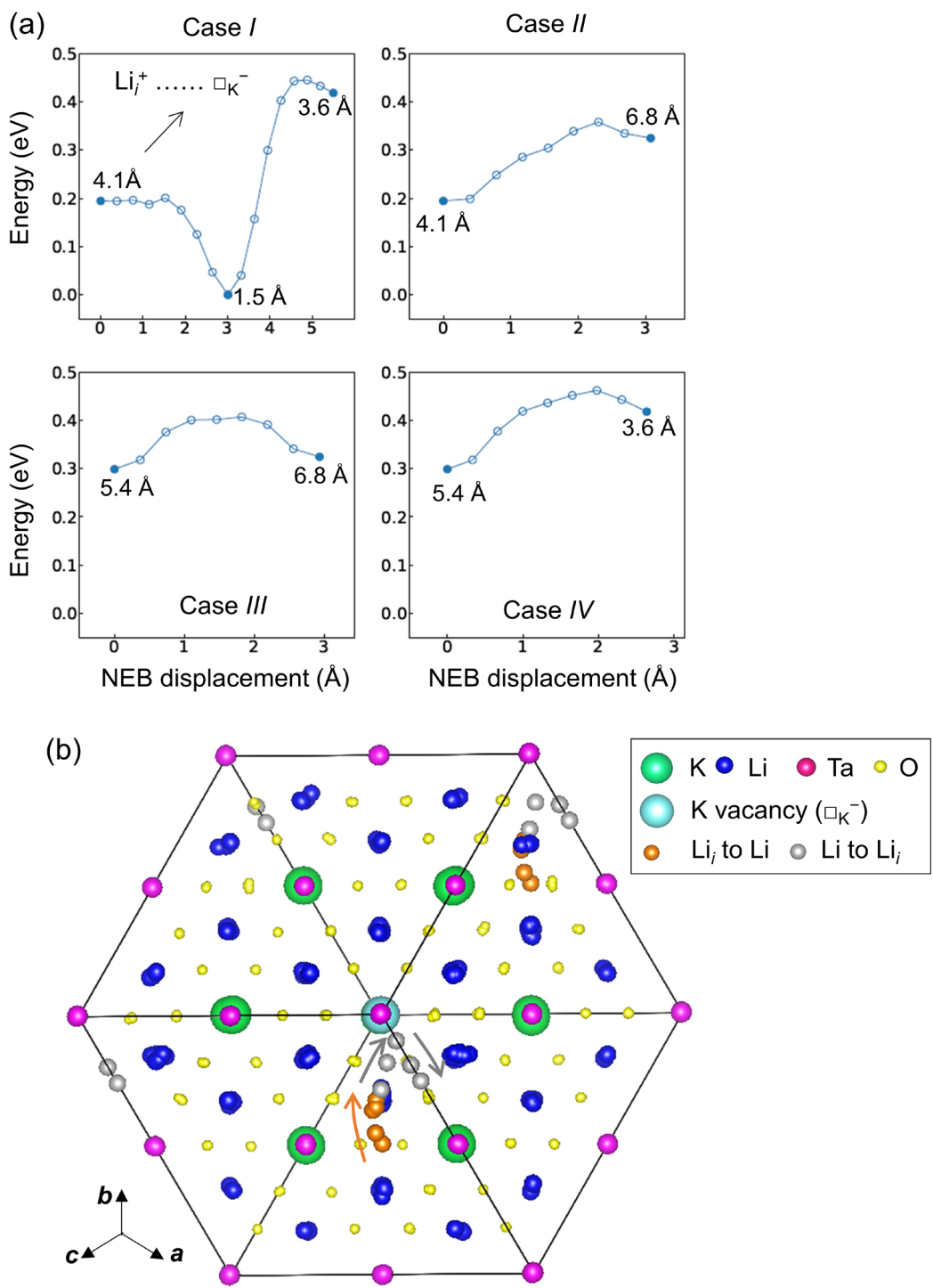

Figure S3. (a) NEB profiles of $Li_i^+$ with different distances from $\square_K^-$ in Li-excess and K-deficient $KLi_6TaO_6$ (using a 2 × 2 × 2 primitive cell with the formula $K_7Li_{49}Ta_8O_{48}$). The numerical values written near the initial or final states indicate the distance between $\square_K^-$ and $Li_i^+$. The energies of the states are shifted with respect to the most stable state where the distance between $\square_K^-$ and $Li_i^+$ is approximately 1.5 Å. The closed and open circles indicate initial and final states where the atomic coordinates were fixed and optimized during the NEB calculations, respectively. (b) The trajectory of the moving atoms during NEB calculations for case *I*. Different from the trajectory shown in Supplementary Figure S2, a moving Li atom escapes from the typical migration path, and is trapped at the $\square_K^-$ site.

Table S1. The ionic radius $(R)$[1,2] and $E_m$ values of 34 types of $A\text{Li}_6X\text{O}_6$. These oxides could exhibit reasonable $E_m$ values obtained by the migration path shown in Figure 2a under Li-excess conditions.

| $A$ | $X$ | Valence of $A$ | Valence of $X$ | $R_A$ (Å)[a] | $R_X$ (Å)[b] | $R_A - R_X$ (Å) | $E_m$ (eV) | Dynamical stability by imaginary phonon frequency |
|---|---|---|---|---|---|---|---|---|
| Ag | V | 1 | 5 | 1.28 | 0.54 | 0.74 | 0.20 | Unstable |
| Ag | Nb | 1 | 5 | 1.28 | 0.64 | 0.64 | 0.29 | Unstable |
| Ag | Ta | 1 | 5 | 1.28 | 0.64 | 0.64 | 0.28 | Unstable |
| Ag | As | 1 | 5 | 1.28 | 0.46 | 0.82 | 0.02 | Unstable |
| Ag | Sb | 1 | 5 | 1.28 | 0.60 | 0.68 | 0.21 | Unstable |
| Ag | Bi | 1 | 5 | 1.28 | 0.76 | 0.52 | 0.29 | Unstable |
| Na | V | 1 | 5 | 1.39 | 0.54 | 0.85 | 0.15 | Unstable |
| Na | Nb | 1 | 5 | 1.39 | 0.64 | 0.75 | 0.23 | Unstable |
| Na | Ta | 1 | 5 | 1.39 | 0.64 | 0.75 | 0.22 | Unstable |
| Na | As | 1 | 5 | 1.39 | 0.46 | 0.93 | 0.05 | Unstable |
| Na | Sb | 1 | 5 | 1.39 | 0.60 | 0.79 | 0.18 | Unstable |
| Na | Bi | 1 | 5 | 1.39 | 0.76 | 0.63 | 0.25 | Unstable |
| Sr | Ti | 2 | 4 | 1.44 | 0.61 | 0.84 | 0.07 | Unstable |
| Sr | Zr | 2 | 4 | 1.44 | 0.72 | 0.72 | 0.17 | Unstable |
| Sr | Hf | 2 | 4 | 1.44 | 0.71 | 0.73 | 0.15 | Unstable |
| Sr | Ge | 2 | 4 | 1.44 | 0.53 | 0.91 | 0.05 | Unstable |
| Sr | Sn | 2 | 4 | 1.44 | 0.69 | 0.75 | 0.12 | Unstable |
| Ba | Ti | 2 | 4 | 1.61 | 0.61 | 1.01 | 0.05 | Unstable |
| Ba | Zr | 2 | 4 | 1.61 | 0.72 | 0.89 | 0.15 | Unstable |
| Ba | Hf | 2 | 4 | 1.61 | 0.71 | 0.90 | 0.14 | Unstable |
| Ba | Ge | 2 | 4 | 1.61 | 0.53 | 1.08 | 0.05 | Unstable |
| Ba | Sn | 2 | 4 | 1.61 | 0.69 | 0.92 | 0.11 | Unstable |
| K | V | 1 | 5 | 1.64 | 0.54 | 1.10 | 0.04 | Stable |
| K | Nb | 1 | 5 | 1.64 | 0.64 | 1.00 | 0.08 | Stable |
| K | Ta | 1 | 5 | 1.64 | 0.64 | 1.00 | 0.09 | Stable |
| K | As | 1 | 5 | 1.64 | 0.46 | 1.18 | 0.06 | Stable |
| K | Sb | 1 | 5 | 1.64 | 0.60 | 1.04 | 0.08 | Stable |
| K | Bi | 1 | 5 | 1.64 | 0.76 | 0.88 | 0.13 | Stable |
| Rb | V | 1 | 5 | 1.83 | 0.54 | 1.29 | 0.05 | Stable |
| Rb | Nb | 1 | 5 | 1.83 | 0.64 | 1.19 | 0.07 | Stable |
| Rb | Ta | 1 | 5 | 1.83 | 0.64 | 1.19 | 0.08 | Stable |
| Rb | As | 1 | 5 | 1.83 | 0.46 | 1.37 | 0.07 | Stable |
| Rb | Sb | 1 | 5 | 1.83 | 0.60 | 1.23 | 0.06 | Stable |
| Rb | Bi | 1 | 5 | 1.83 | 0.76 | 1.07 | 0.12 | Stable |

[a] $R_A$ values were obtained from a coordination number of 12. Ag values were obtained from a coordination number of 8.

[b] $R_X$ values were obtained from coordination number of 6.

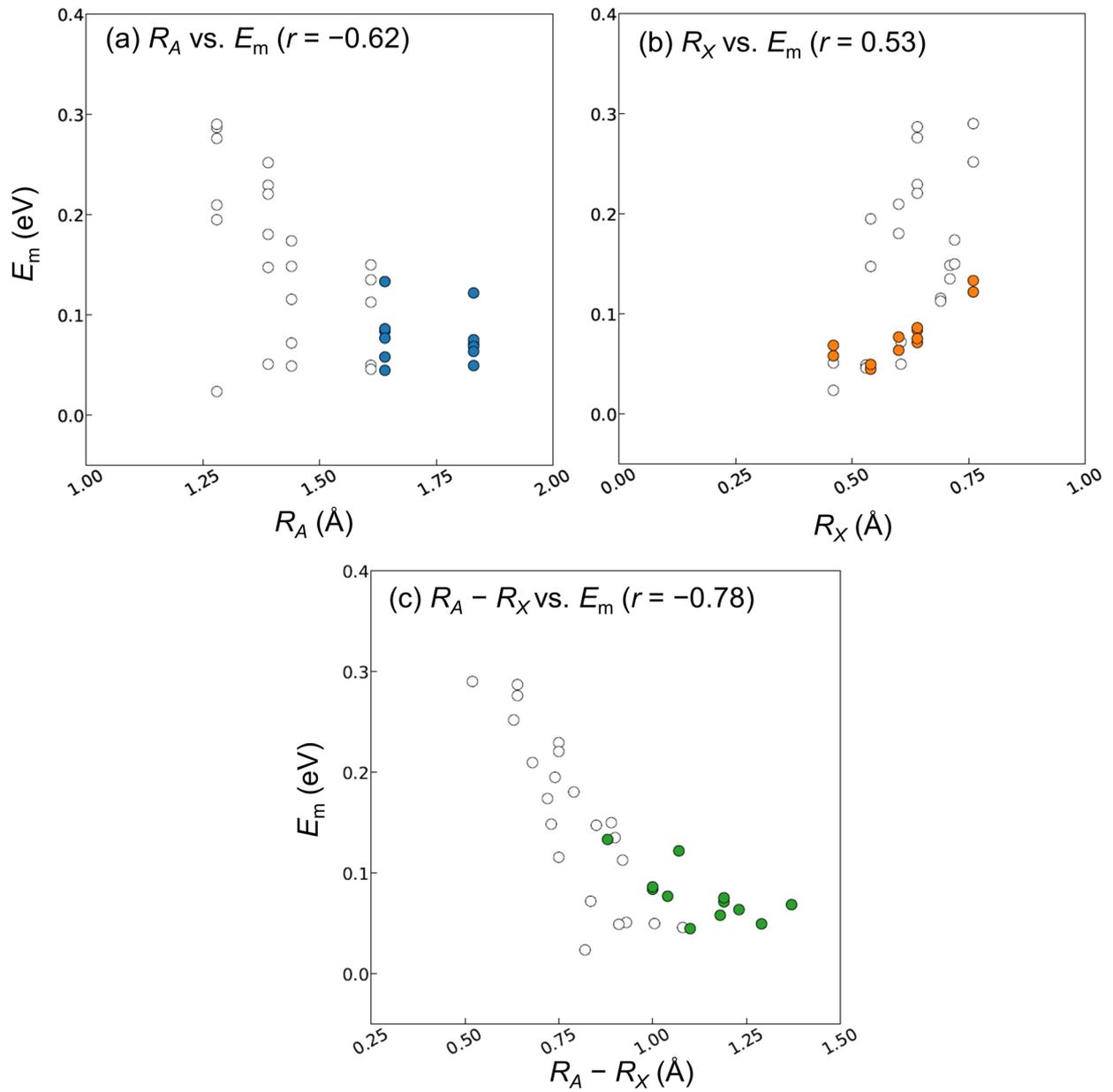

Figure S4. $E_m$ dependence on (a) ionic radius of $A$ ($R_A$),[1,2] (b) ionic radius of $X$ ($R_X$), and (C) $R_A - R_X$ of 34 types of $A$Li$_6$$X$O$_6$ listed in Supplementary Table S1. Colored closed marks indicate the 12 suggested $A$Li$_6$$X$O$_6$ that are dynamically stable. Open marks indicate the dynamically unstable oxides identified by imaginary phonon frequencies. $r$ is a linear correlation coefficient.

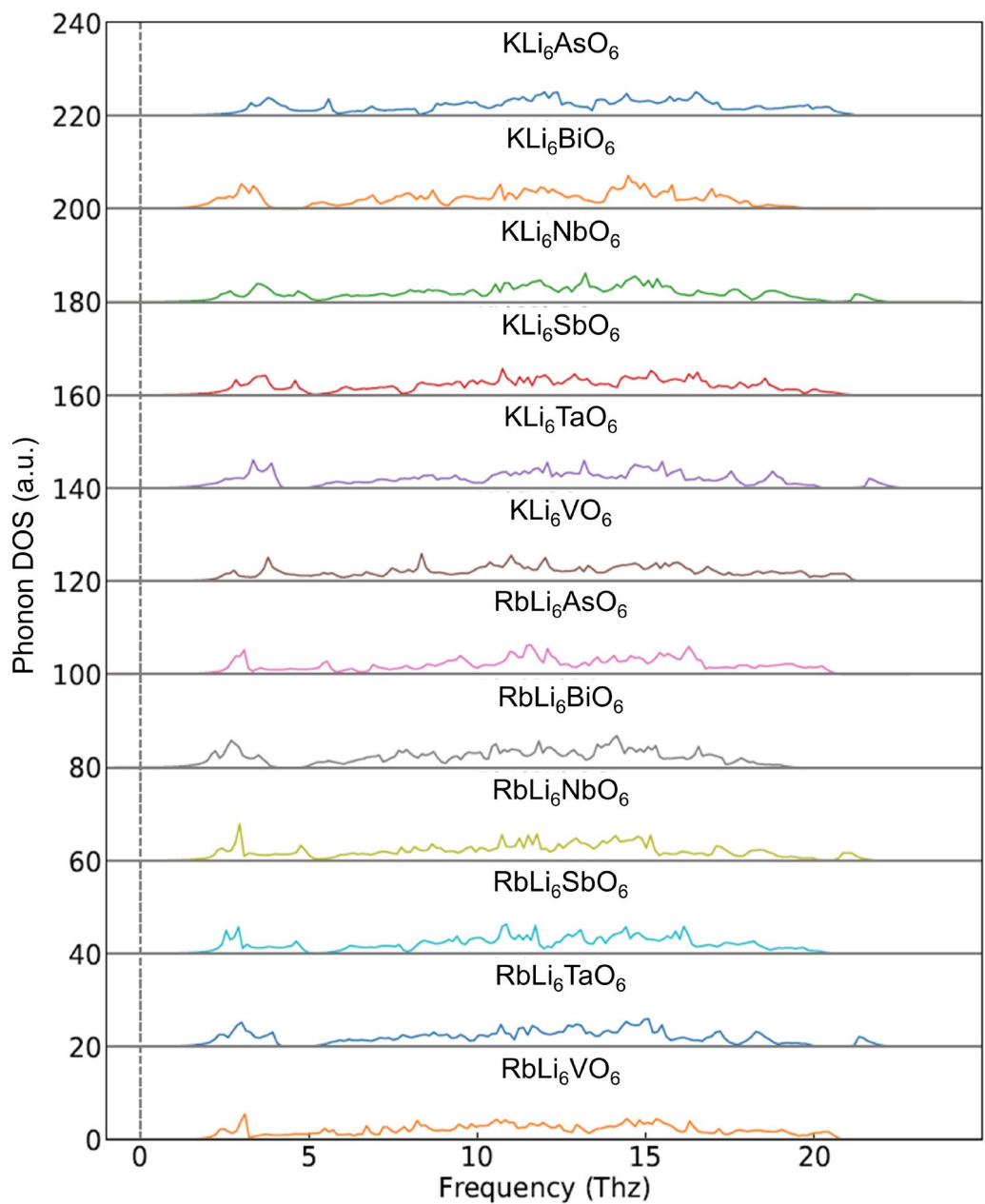

Figure S5. Phonon density-of-states of the 12 suggested $A$Li$_6$$X$O$_6$ structures. The absence of phonon density-of-states at the negative frequencies indicates that their crystal structures are dynamically stable.



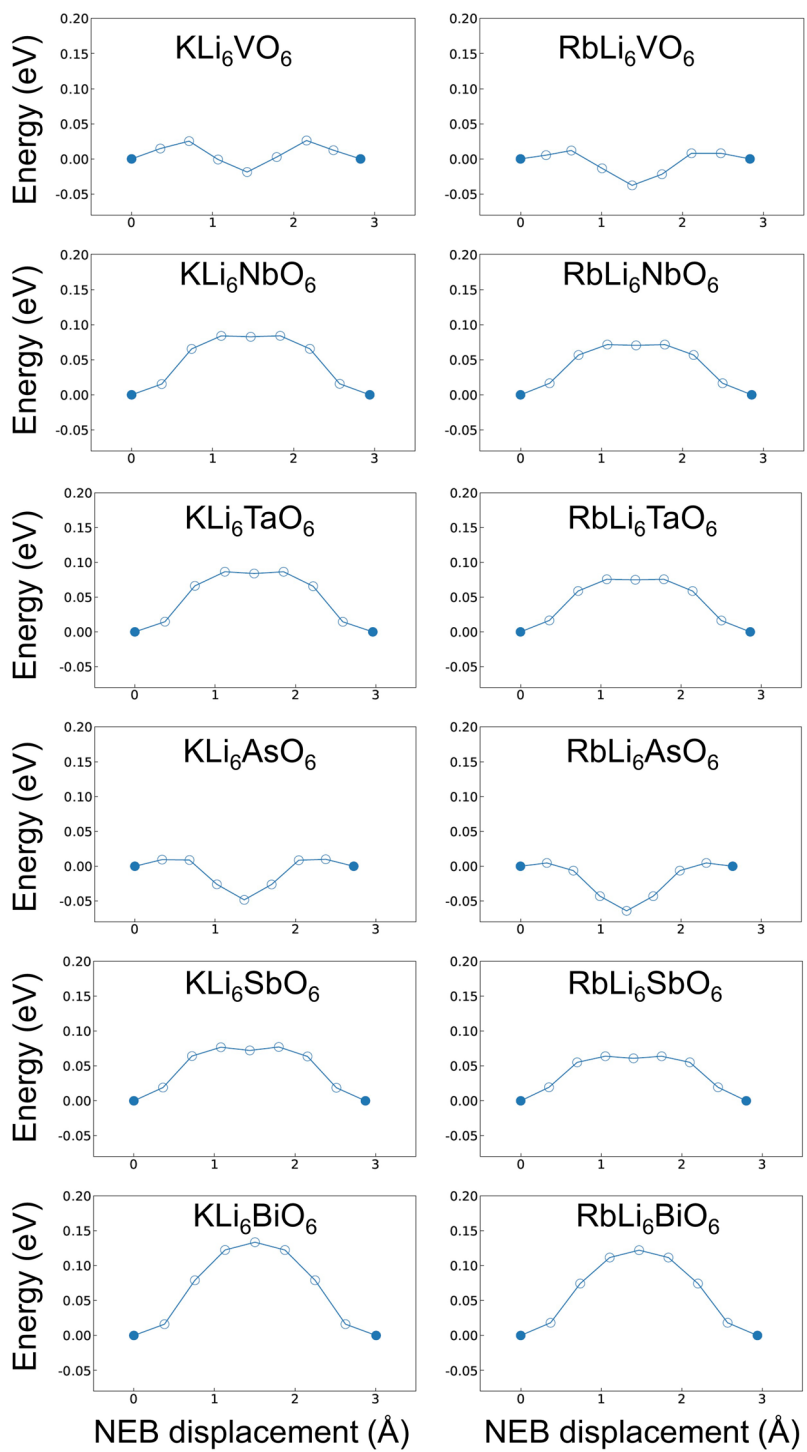

Figure S6. NEB profiles of $Li_i^+$ in the suggested $A$Li$_6$$X$O$_6$ structures (using a 2 × 2 × 2 primitive cell with the formula $A_8$Li$_{49}$$X_8$O$_{48}$). The closed and open circles indicate initial and final states where the atomic coordinates were fixed and optimized during the NEB calculations, respectively.

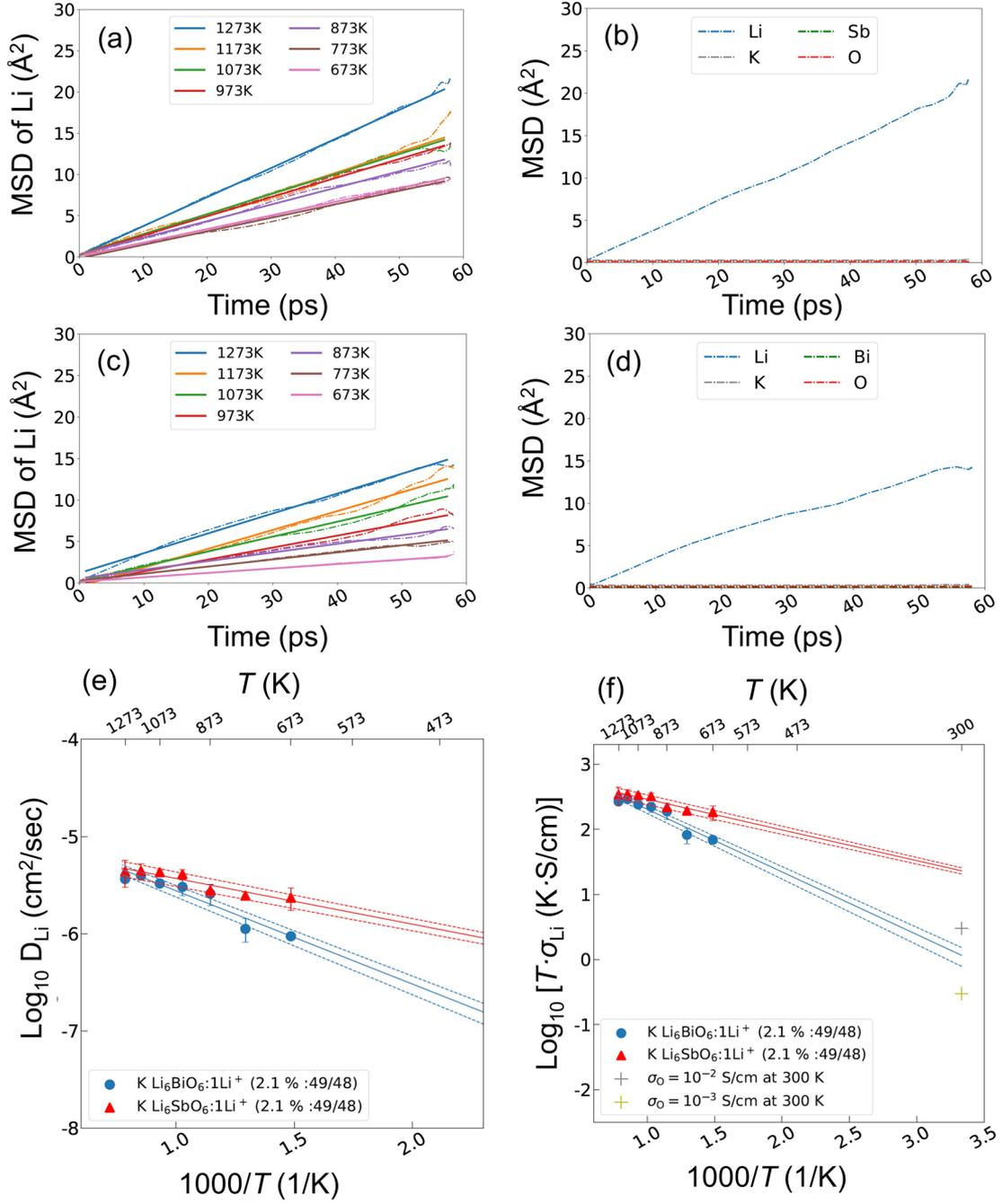

Figure S7. The relationship between simulation time and mean-squared-displacement (MSD) of (a) Li atoms at multiple temperatures and (b) constituent elements at 1273 K in the Li-excess KLi$_6$SbO$_6$ showed the lowest $E_a$ among the 12 suggested $A$Li$_6X$O$_6$. The relationship between the simulation time and MSD of (c) Li atoms at multiple temperatures and (d) constituent elements at 1273 K in the Li-excess KLi$_6$BiO$_6$ showed the highest $E_a$. The dotted lines indicate MSD obtained by using Eq. 8 and the solid lines indicate the fitted line. The MSD curve for Li atoms and the fitted line showed a linear correlation with $r > 0.97$ for the 12 suggested $A$Li$_6X$O$_6$ at all the simulation temperatures. For all FPMD simulations, the elements other than Li showed a significantly small MSD < 1 Å$^2$. Dependence of (e) $D_{Li}$ and (f) temperature × $\sigma_{Li}$ on the inverse temperature in the Li-excess KLi$_6$SbO$_6$ and KLi$_6$BiO$_6$. The error bar is obtained from ± one standard deviation from three different FPMD trials. The solid lines indicate the estimated $D_{Li}$ and $\sigma_{Li}$ by using Eqs. 9 and 10, respectively. The dashed line indicates the estimated values from the maximum and minimum points within the error range. The estimated $D_{Li}$ and the values obtained by FPMD simulations showed a linear correlation with $r > 0.89$ for the 12 suggested $A$Li$_6X$O$_6$.

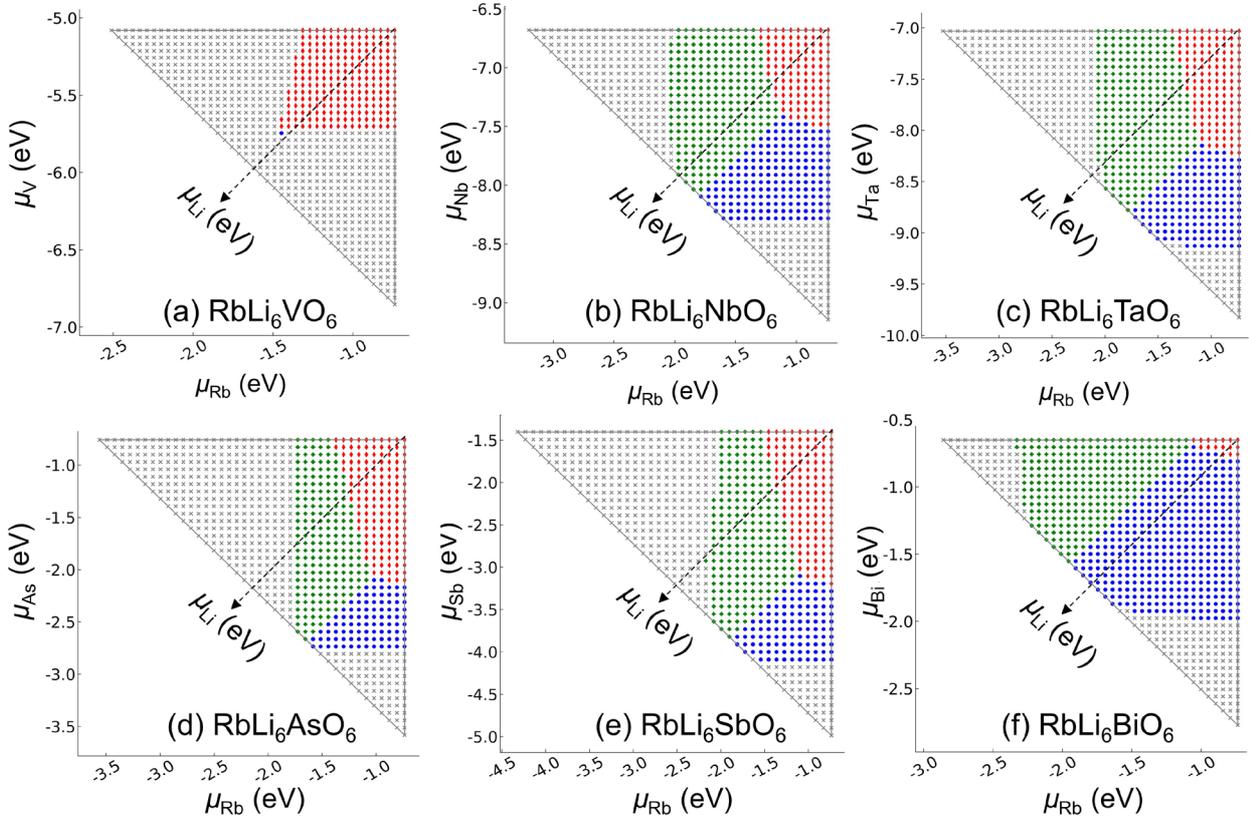

Figure S8. The stability region (triangle) of the suggested RbLi$_6$XO$_6$ structures, (a) RbLi$_6$VO$_6$, (b) RbLi$_6$NbO$_6$, (c) RbLi$_6$TaO$_6$, (d) RbLi$_6$AsO$_6$, (e) RbLi$_6$SbO$_6$, and (f) RbLi$_6$BiO$_6$ in $\mu_{Rb}$ and $\mu_X$ at a temperature of 1073 K with limits imposed by binary oxides. Colored symbols inside the triangle indicate the condition relevant to the concentration and $E_f$ of point defects as variables of $\mu$. The calculations for $E_f$ were obtained by GGA parameterized in the PBE form. The detailed information for each symbol can be referred to in Figure 3a.

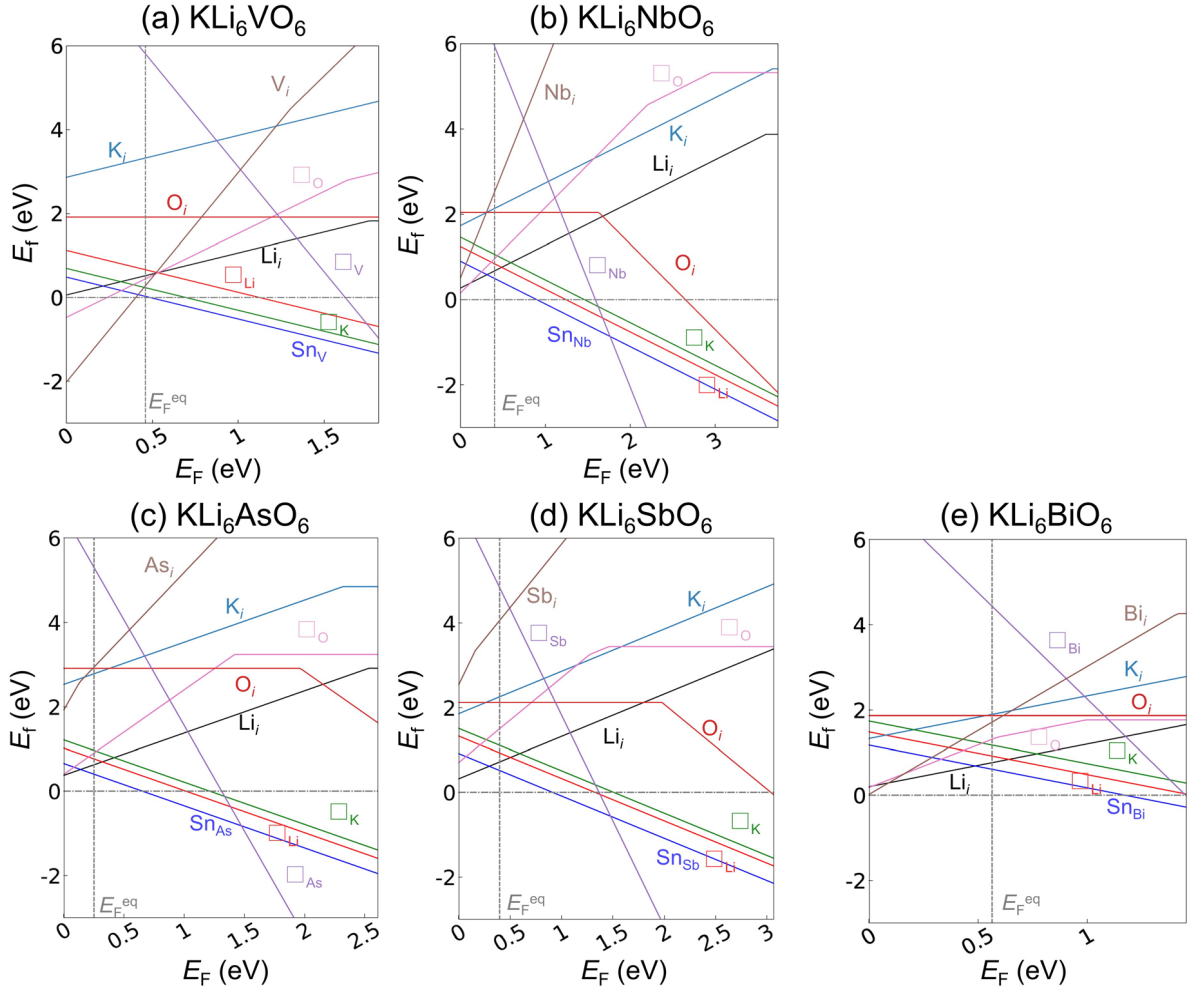

Figure S9. $E_f$ of native defects and a dopant of (a) $KLi_6VO_6$, (b) $KLi_6NbO_6$, (c) $KLi_6AsO_6$, (d) $KLi_6SbO_6$, (e) $KLi_6BiO_6$ as a variable of $E_F$. $\mu$ is located at the position presented as a symbol * in Figure 5. The calculations were obtained by GGA parameterized in the PBE form. The gradient indicates the charge of a defect. The lowest $E_f$ obtained among various charges is only displayed for each defect. $E_F^{eq}$ is an equilibrium Fermi level that is calculated by the charge neutrality condition.

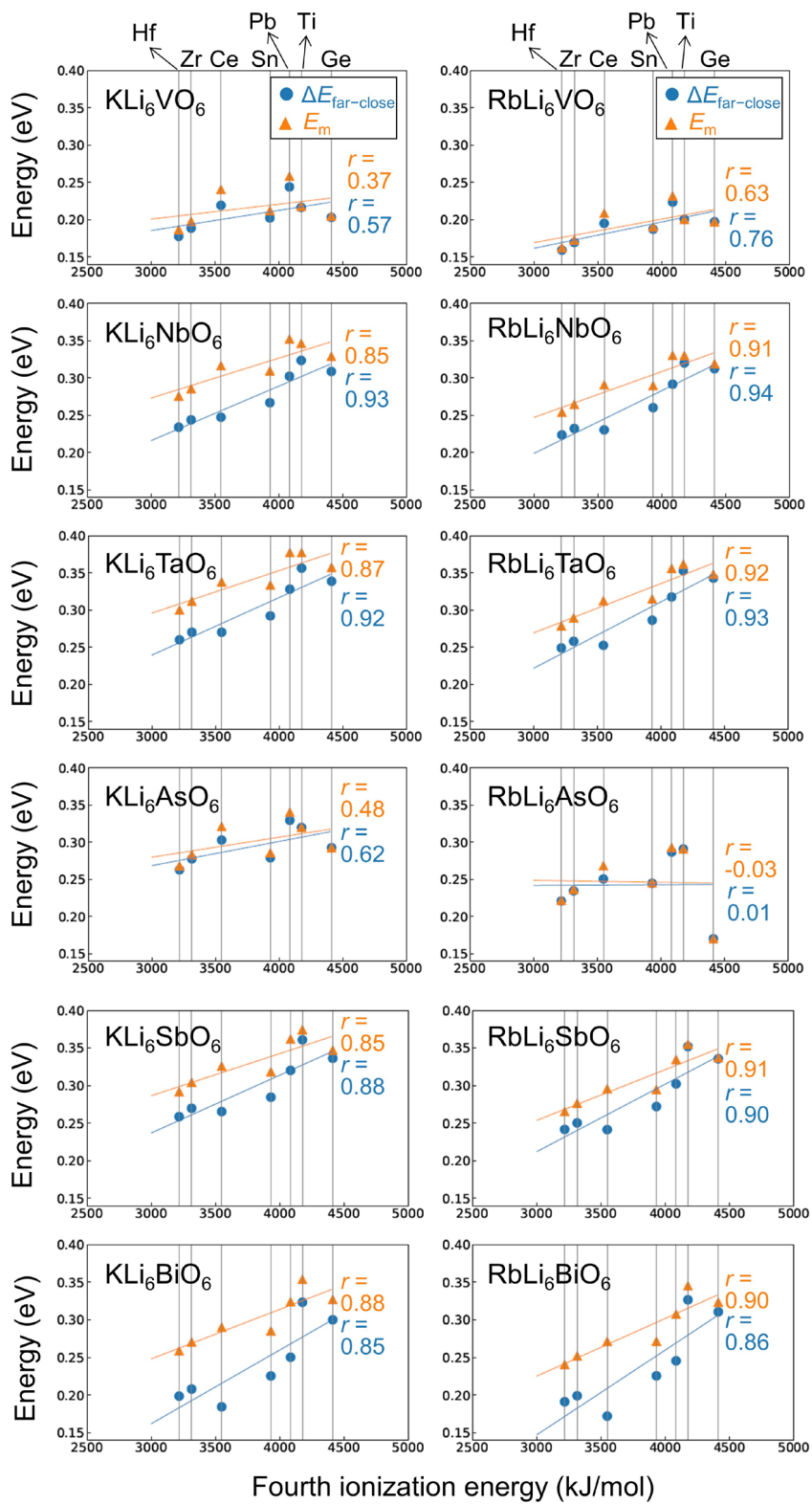

Figure S10. Relationships between the fourth ionization energy of dopant $M$ (vertical gray lines) and $E_m$ (orange triangles) or $\Delta E_{\text{far–close}}$ (blue circles) of the suggested (K or Rb)Li$_6$XO$_6$ with substitution of $M$ for $X$. Colored lines were obtained *via* ordinary linear regression, and $r$ indicates the linear correlation coefficient.

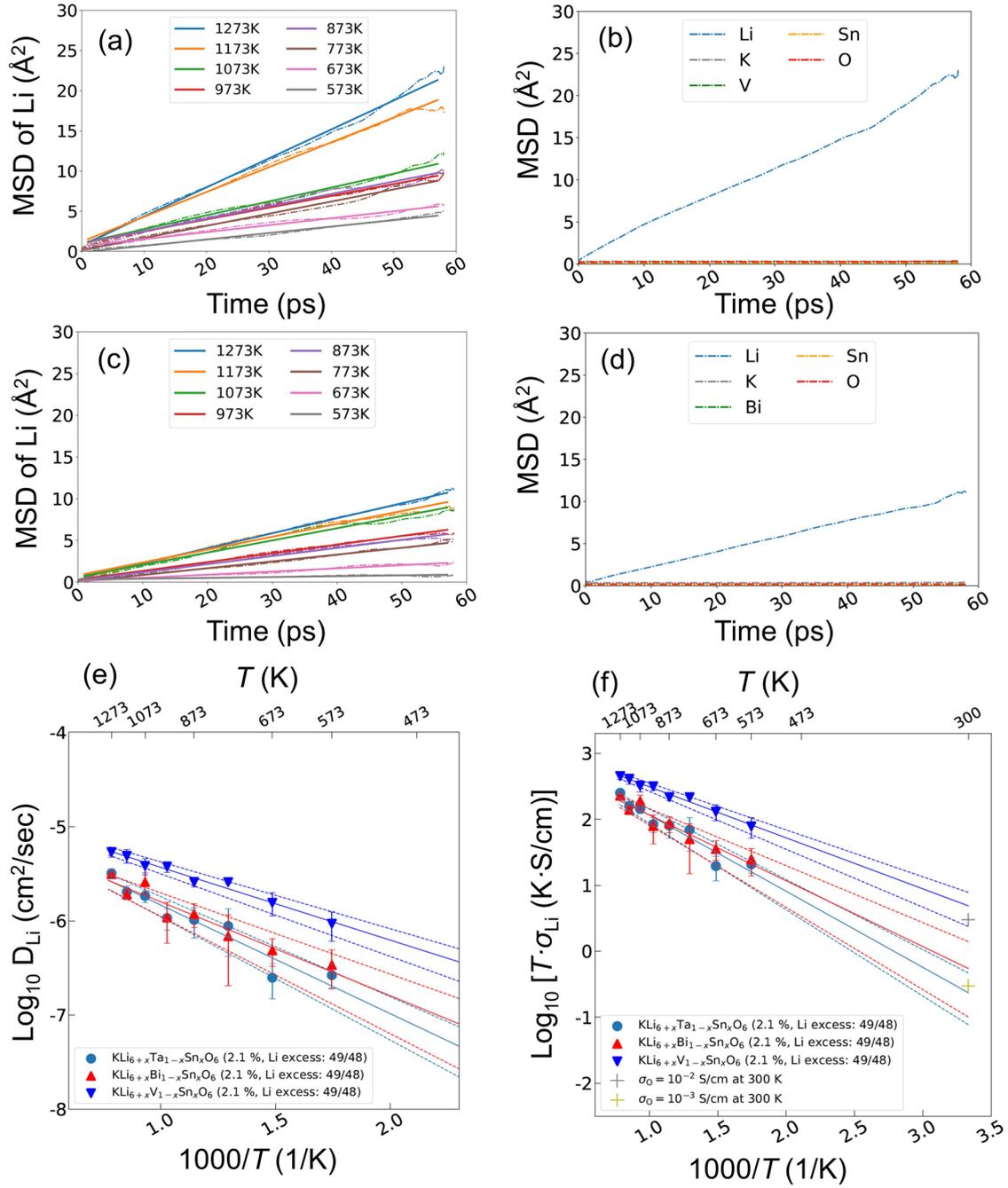

Figure S11. Relationship between the simulation time and mean-squared-displacement (MSD) of (a) Li atoms at multiple temperatures and (b) constituent elements at 1273 K in the Sn-doped $KLi_6VO_6$ (2 × 2 × 2 primitive cell of the formula $K_8Li_{49}SnV_7O_{48}$) under Li-excess conditions. Relationship between the simulation time and MSD of (c) Li atoms at multiple temperatures and (d) constituent elements at 1273 K in the Sn-doped $KLi_6BiO_6$ under Li-excess conditions. The dotted lines indicate MSD obtained by using Eq. 8 and the solid lines indicate the fitted line. Dependence of (e) $D_{Li}$ and (f) temperature × $\sigma_{Li}$ on the inverse temperature in the Sn-doped $KLi_6VO_6$, $KLi_6BiO_6$, and $KLi_6TaO_6$ under Li-excess conditions. The error bar is obtained from ± one standard deviation from three different FPMD trials. The solid lines indicate the estimated $D_{Li}$ and $\sigma_{Li}$ by using Eqs. 9 and 10, respectively. The dashed line indicates the estimated values from the maximum and minimum points within the error range.

**Supplementary References**